\documentclass[aps,prb,reprint,twocolumn,superscriptaddress,floatfix,nofootinbib]{revtex4-2}
\usepackage{amsmath,amssymb,color,comment,physics}
\usepackage[makeroom]{cancel}
\usepackage{makecell}
\usepackage{bbm}
\usepackage[caption=false]{subfig}
\usepackage{mathrsfs}
\usepackage{graphicx}
\usepackage{subfig}
\usepackage[countmax]{subfloat}
\usepackage[english]{babel}
 \usepackage{dsfont}
 \usepackage{hyperref}
\hypersetup{colorlinks,linkcolor=OrangeRed,urlcolor=NavyBlue,citecolor=RoyalBlue}
\usepackage[dvipsnames]{xcolor}
\usepackage{braket}
\usepackage{color}

\definecolor{mypurple}{rgb}{0.49,0.18,0.56}
\definecolor{mygold}{rgb}{0.93,0.59,0.13}
\definecolor{mygreen}{rgb}{0,0.5,0}
\definecolor{myblue}{rgb}{0,0,0.75}
\definecolor{mymagenta}{cmyk}{0,1,0,0.12}
\definecolor{mygray}{rgb}{0.5,0.5,0.5}

\usepackage{umoline}

\newcommand{\Sx}[1]{\sigma^{x}_{#1}}
\newcommand{\Sy}[1]{\sigma^{y}_{#1}}
\newcommand{\Sz}[1]{\sigma^{z}_{#1}}
\newcommand{\SP}[1]{S^{+}_{#1}}
\newcommand{\SM}[1]{S^{-}_{#1}}
\newcommand{\SX}[1]{S^{x}_{#1}}

\newcommand{\SZ}[1]{S^{z}_{#1}}

\begin{document}

\title{Towards the real-time evolution of gauge-invariant $\mathbb{Z}_2$ and $U(1)$ quantum link models on NISQ Hardware with error-mitigation}
%\title{Real-time dynamics of Plaquette Models using NISQ Hardware}
\author{Emilie Huffman}
\affiliation{Perimeter Institute For Theoretical Physics, Waterloo, Canada}
\author{Miguel Garc\'ia Vera}
\affiliation{Departamento de F\'isica, Escuela Polit\'ecnica Nacional, Av. Ladr\'on de Guevara E11-253, Quito, Ecuador}
\author{Debasish Banerjee}
\affiliation{Saha Institute of Nuclear Physics, HBNI, 1/AF Bidhannagar, Kolkata 700064, India}
\affiliation{Homi Bhabha National Institute, Training School Complex, Anushaktinagar, Mumbai 400094, India}

\begin{abstract}
 Practical quantum computing holds clear promise in addressing problems not generally 
 tractable with classical simulation techniques, and some key physically interesting 
 applications are those of real-time dynamics in strongly coupled lattice gauge theories. 
 In this article, we benchmark the real-time dynamics of $\mathbb{Z}_2$ and $U(1)$ gauge 
 invariant plaquette models using noisy intermediate scale quantum (NISQ) hardware, 
 specifically the superconducting-qubit-based quantum IBM Q computers. We design quantum 
 circuits for models of increasing complexity and measure physical observables such as 
 the return probability to the initial state, and locally conserved charges. NISQ hardware 
 suffers from significant decoherence and corresponding difficulty to interpret the results. 
 We demonstrate the use of hardware-agnostic error mitigation techniques, such as circuit 
 folding methods implemented via the Mitiq package, and show what they can achieve within 
 the quantum volume restrictions for the hardware. Our study provides insight into the 
 choice of Hamiltonians, construction of circuits, and the utility of error mitigation 
 methods to devise large-scale quantum computation strategies for lattice gauge theories.

\end{abstract}
\date{\today}
\maketitle
\tableofcontents
\section{Introduction}
 Gauge theories are a cornerstone in the description of various naturally
occurring phenomena in Nature, whether in particle or in condensed matter
physics \cite{Wilczek_2016}. These theories are characterized by the presence
of local conservation laws, which are in general not enough to make the models
integrable. However, such local conservation laws greatly constrain these systems,
leading to exotic phenomena involving quantum  entanglement of the fundamental
degrees of freedom over long distances, many of which remain unexplored due to
computational difficulties to study them on a classical computer. In addition,
one of the outstanding challenges in fundamental physics is to study real-time
dynamics of the quantum entanglement inherent in gauge theories that leads to
confinement. The rapid experimental development of quantum computers (both
analog and digital) \cite{Preskill2018quantumcomputingin, Noh_2016,superconductingQubits, Lanyon_2011, bloch_quantum_2012} 
following the pioneering suggestion of Feynman \cite{feynman_simulating_nodate} 
provides an opportunity to overcome these bottlenecks and make new fundamental 
progress in this field.

 While certain initial exciting developments have been obtained from the studies 
of finite, relatively small systems using classical computations such as exact
diagonalization and variational methods using the MPS ans\"{a}tze, it is pertinent
to understand the corresponding behaviour in large quantum systems. This is an
exponentially difficult problem in the system size for most of the classical
computational methods in use, thus demanding the use of new toolboxes 
such as quantum computers. Although theoretically promising, current quantum
computers in use are either of the analog variety, where a certain experimental 
set-up can very efficiently emulate only a limited variety of physical systems; 
or of the digital kind, which are limited by the moderate number of 
available (noisy) qubits. There has however, been some progress towards the
development of hybrid analog-digital approaches with the aim to combine the 
desirable features of both \cite{Parra_Rodriguez_2020}. For the case of
digital quantum computation, which will be our main focus in this article, it
becomes important to devise efficient optimizations of the quantum circuitry so that
the studies can be extended to large quantum systems. The results need to be
benchmarked from an independent computational method at small or medium system
sizes. While such studies have been extensively carried out for spin
models, implementations of quantum link models on quantum hardware are relatively 
scarce, a gap which our article aims to fill.

 Moreover, one of the crucial theoretical physics problems where quantum
computers could play a central role is establishing the emergence of
thermalization in isolated many-body quantum systems, necessary to describe
equilibrium properties of the system using quantum statistical mechanics
\cite{PhysRevE.50.888, PhysRevA.43.2046}. This has become well-known in the
literature under the eigenstate thermalization hypothesis (ETH). On the other hand,
in the absence of thermalization, the properties of the initial states are
preserved for a long time, and the growth of quantum entanglement is very slow.
This is known to occur in the many-body localized (MBL) phases \cite{ALET2018498},
and has raised the possibility of using such phases as quantum memories, which
can encode quantum information with high fidelity \cite{2016NatPh12907S}.
Confining phases of gauge theories could potentially offer the possibility of 
realizing topologically stable qubits, unaffected by local decoherent noise,
and act as quantum memories. Another relatively new development is the discovery of
atypical quantum states in (strongly) interacting quantum systems, dubbed as quantum
many body scars \cite{Serbyn_2021}, which do not follow the ETH unlike other quantum
states. Even though such states belong to the highly excited part of the energy spectrum,
they have anomalously low entropy. Studying properties of such quantum states on 
large systems would also benefit from a quantum computer, given the computational
complexity for classical simulation methods.

 In the context of particle physics, especially for non-perturbative ab-initio
computations in lattice chromodynamics (LQCD), a plethora of questions involving
physics at real-time and high baryon density cannot be reliably answered using 
classical algorithms running on classical computers. Quantum computers, both
analog and digital, have been proposed in order to make progress in this front
\cite{Banuls:2019bmf}. Several pioneering experiments \cite{Martinez:2016yna, 
Bernien_2017, Schweizer_2019, Mil:2019pbt, Yang:2020yer, Davoudi:2019bhy} have
already demonstrated the possibility of harnessing the new technology to address
questions posed in the context of high-energy physics (HEP). Further, the 
availability of noisy intermediate-scale (universal) quantum computers from the IBM
and the Rigetti corporations have empowered the theorists to perform experiments.
Recently, there have been many such preliminary efforts to address representative
questions in simpler gauge theories using quantum computing techniques. These
include investigation of scattering and real-time dynamics in spin systems 
\cite{Lamm_2018, Gustafson:2019mpk, gustafson2021benchmarking} and in gauge theories
\cite{Klco_2018, Klco_2020}, static charges in gauge theories \cite{Zhang_2018}, as
well as mass spectra in Abelian and non-Abelian lattice gauge theories 
\cite{Lewis:2019wfx, Atas:2021ext}. Naturally, the efforts to represent only 
physical states of the corresponding gauge theory Hamiltonian, which are
invariant under the Gauss law, in the limited quantum hardware available to us
have spurred a  cascade of theoretical developments \cite{Stryker:2018efp,Raychowdhury:2019iki, Raychowdhury:2018osk, Davoudi:2020yln, Klco:2018zqz,Klco:2020aud, Ciavarella_2021, Bender_2020, aidelsburger2021cold,zohar2021quantum, kasper2020universal,Funcke_2021}. 

  A major obstacle in the design of quantum circuits and quantum algorithms is 
the decoherence of the superconducting qubits in contemporary quantum computers, 
also called noisy intermediate scale quantum (NISQ) devices, such as the IBM Q
and the Rigetti platforms. The qubits in these devices are only approximately 
isolated from the environment, and the gate operations needed to induce some 
interaction terms among them also depend on whether the operation is a single, or 
a multi-qubit operation (the latter have smaller fidelities). Moreover, single gate 
operations can have different gate times depending on the specific qubit they are 
applied to. These factors induce errors in the measured quantities, and although 
quantum error correction schemes have been devised decades ago 
\cite{PhysRevA.52.R2493, PhysRevLett.77.793}, their implementation is hindered by 
the fact that they require additional qubits to correct the error on a single qubit, 
making them impractical for NISQ era devices with a limited number of available
qubits (typically of the order of 6-10). A recent alternate approach exploits 
the available qubits, but repeats the experiments for a different number of times, 
and with different sets of quantum gates. The resulting data can be extrapolated 
to the case when there is no noise affecting the experiment, assuming a general 
noise model. This approach, known as the zero noise extrapolation (ZNE) and has been 
intensively investigated in 
\cite{PhysRevX.7.021050, Kandala_2019,He_2020, larose2020mitiq, Giurgica_Tiron_2020, lowe2020unified, sopena2021simulating}. 
It falls into the category of error mitigation rather than error correction. 
Schemes for addressing depolarizing errors have been investigated in \cite{PhysRevLett.122.180501}, 
and readout errors in \cite{funcke2020measurement, Nachman2020, Jattana2020}. 
Proposals of correcting depolarizing noise in a hierarchical fashion in quantum 
circuits depending on whether they contribute to the UV or IR physics have been 
put forward in \cite{klco2021hierarchical}, and would allow targeted improvements 
in scientific applications in appropriate energy windows.

 Our main goal in this article is to present models and implement corresponding 
quantum circuits suitable for NISQ devices for simulating real-time dynamics in pure 
gauge theories on single and double plaquettes. The plaquette interaction has been
considered before in \cite{Lewis:2019wfx} following the usual Wilson formulation of
formulating lattice gauge fields, having an infinite dimensional Hilbert space for each
link degree of freedom.  This necessarily needs a truncation in the allowed set of
states to be represented in an architecture with a finite number of qubits. Instead, we
will consider a different formulation of lattice gauge theories, which are commonly
known as quantum link models (QLMs) 
\cite{Horn:1981kk, Orland:1989st,Chandrasekharan:1996ih}.
This formulation is ideally suited for implementation in quantum computers, since
gauge invariance is realized exactly with a finite-dimensional Hilbert space for
each link degree of freedom. In fact, the dimensionality of the local Hilbert space
can be tuned in a gauge-invariant manner.

The strength of QLMs for NISQ devices is illustrated quantitatively in Table 
\ref{datatable2} (see Supplementary Material Appendix C for more details), where the 
minimum number of two-qubit gates needed per qubit to simulate a single Trotter step 
of the time-evolution of gauge theory potential terms is given for QLMs as well as 
truncated Wilson theories. A $d$-dimensional square lattice is assumed, and the circuit 
implementation used is the one we use in our simulations, and is described in Section III. 
The Wilson column refers to the potential terms of the Kogut-Susskind Hamiltonian 
\cite{Carena:2022kpg,PhysRevD.103.114505}, and the Improved Wilson column is for the 
Symanzik correction terms which have been proposed to reduce the number of Trotter steps 
necessary for a simulation \cite{Carena:2022kpg}. While the Kogut-Susskind Hamiltonian 
and Symanzik improvement have the prospect of being very useful for simulating gauge 
theories in the future of quantum computing, Table \ref{datatable2} makes it clear that 
quantum link models are much more suited for taking the first steps of simulating 
time-evolution for gauge theories on real hardware, with the aforementioned advantage 
of being gauge-invariant at every tuning step. In fact, even exactly gauge-invariant QLMs 
of non-Abelian theories are in much closer reach for time-evolution than alternative 
formulations, for example an $SO(3)$-symmetric theory would require $162 (2d-2)$ 
two-qubit gates per qubit per Trotter step \cite{Chandrasekharan:1996ih,Rico_2018}.

\begin{table}[htb]
\begin{center}
\small
\begin{tabular}{|l|l|l|l|}
\hline
\makecell[l]{\textbf{Gauge}\\ \textbf{Group}} & \textbf{QLM} & \textbf{Wilson} & \makecell[l]{\textbf{Improved} \\\textbf{Wilson} } \\
 \hline
$\mathbb{Z}_2$ & $2(2d-2)$ & $2(2d-2)$ & \makecell[l]{$2\cdot 3(2d-2)$\\$+2(2d-4)(2d-2)$}\\
\hline
$U(1)$ & $16(2d-2)$ & $2\cdot 2048 (2d-2)$ & \makecell[l]{$2\cdot 32\cdot 4096$\\$ \;\cdot 3 (2d-2)$\\+$2\cdot 32 \cdot 4096 $\\$\; \cdot (2d-4)(2d-2)$}\\
\hline
\end{tabular}
\vspace{.5cm}
\caption{The number of two-qubit gates necessary for each qubit that corresponds to a link, 
for a single Trotter step and as a function of square lattice dimension $d$. Details are in 
Supplementary Material Appendix C.}
\label{datatable2}
\end{center}
\end{table}

QLMs are quite popular for
implementation on analog quantum simulators \cite{Bernien_2017, Mil:2019pbt, Yang:2020yer}, 
and it makes sense to develop the corresponding implementation in digital platforms as well. 
Initial studies of construction of quantum circuits for the plaquettes using the QLM approach 
were reported in \cite{2011NJPh13h5007M, Mezzacapo:2015bra}. We focus on the theories with
$\mathbb{Z}_2$ and $U(1)$ local symmetries and explore their formulations on triangular and
square lattice geometries. The Hamiltonians with these local symmetries have been used to 
describe physical systems in condensed matter and quantum information 
\cite{Kitaev_2003,PhysRevB.69.220403, Hermele_2004}.  A quantum circuit for a triangular $U(1)$ 
quantum link model has been proposed in \cite{brower2020lattice} and tested with classical 
hardware. Another recent work dealing with the triangular $U(1)$ quantum link model used 
dualization to obtain dual quantum height variables, which allows a denser encoding in terms 
of qubits \cite{banerjee2021nematic}. To the best of our knowledge, our article is the first 
to demonstrate a hardware-independent error mitigation technique for real-time evolution of 
quantum link lattice gauge theories.

  The rest of the paper is organized as follows. In Section \ref{sec:models} we describe 
the Hamiltonians as well as the corresponding local unitary Abelian transformations which 
keep the Hamiltonian invariant, showing the constrained nature of the Hilbert space in these 
models. In Section \ref{sec:circuits} we describe the quantum circuit used to implement the 
Hamiltonian interactions and perform the real-time dynamics. We outline the methodology we 
adopted in mitigating the errors due to decoherence  and readout in Section \ref{sec:errcorr}; 
and outline the experimental results obtained in Section \ref{sec:results}. Finally, we 
discuss possibilities of extending this study to larger lattice dimensions as well as to 
non-Abelian gauge theories in Section \ref{sec:conc}.
 
 \section{Abelian Lattice Gauge Theory Models} \label{sec:models}
  In this section, we discuss the quantum Hamiltonians, which are invariant under
local $\mathbb{Z}_2$ and the $U(1)$ transformations. The gauge theory Hamiltonians are
characterized by the plaquette term, which is the simplest gauge invariant operator
that can be constructed.

\subsection{The \texorpdfstring{$\mathbb{Z}_2$}{Z(2)} gauge theory}
 Consider a square lattice, for which the smallest closed loop would be a plaquette
containing the four links around an elementary square. Through a four spin
interaction involving $\SZ{} = \Sz{}/2$ operators, and a single spin $\SX{} = \Sx{}/2$
operator on each of the links, we can realize the $\mathbb{Z}_2$ gauge theory Hamiltonian:
 \begin{align}
     H & = -g \sum_{\Box} U_{\Box} - \Gamma \sum_i \SX{i} \, , \\
     U_{\Box} & = \SZ{r,\mu} \SZ{r+\mu,\nu} \SZ{r+\nu,\mu} \SZ{r,\nu} \, .
     \label{z2a}
 \end{align}
The gauge symmetry arises due to the invariance of the Hamiltonian under local
unitary transformations according to the operator:
\begin{equation}
 \begin{aligned}
   V_r &= \Sx{r,\mu} \Sx{r,\nu} \Sx{r-\mu,\mu} \Sx{r-\nu,\nu}\\
   &= \exp \left[i \pi \sum_\mu (\SX{r,\mu} - \SX{r-\mu,\mu})  \right].
 \end{aligned}
 \label{z2b}
 \end{equation}
This can be directly proven from the fact that the Hamiltonian commutes with the
local operator $V_r$, which is known as the Gauss law operator. This commutation
relation $[U_\Box, V_r] = 0$ follows from a few lines of algebra.

The eigenstates of the Hamiltonian are classified into two super-selection sectors
according to $V_r \ket{\psi} = \pm 1 \ket{\psi}$ in the computational basis of
$\Sx{}$. For a square lattice, four links touch a single vertex, and $2^4$
spin configurations are possible, but only half of them have $V_r = 1$ and
the other half $V_r = -1$, giving rise to two super-selection sectors. 

 We are interested in implementing the real-time evolution of simple plaquette
 models on superconducting-qubit-based IBM Q quantum computers. For our purposes,
 we can work in the $\sigma^x$-basis, where the Gauss law as well as the $\Gamma$
 term in the Hamiltonian are diagonal. We aim to start with initial product states
 in the $\Sx{}$ basis, which is then evolved by an off-diagonal plaquette
 Hamiltonian. We note that the $\Gamma$ term not only contributes a diagonal term 
 in this basis but would also be zero for certain Gauss law sectors for the single-plaquette system.
 We choose $\Gamma = 0$ for the experiments performed on the quantum computer.

\begin{figure}
      \centering
      \includegraphics[scale=0.6]{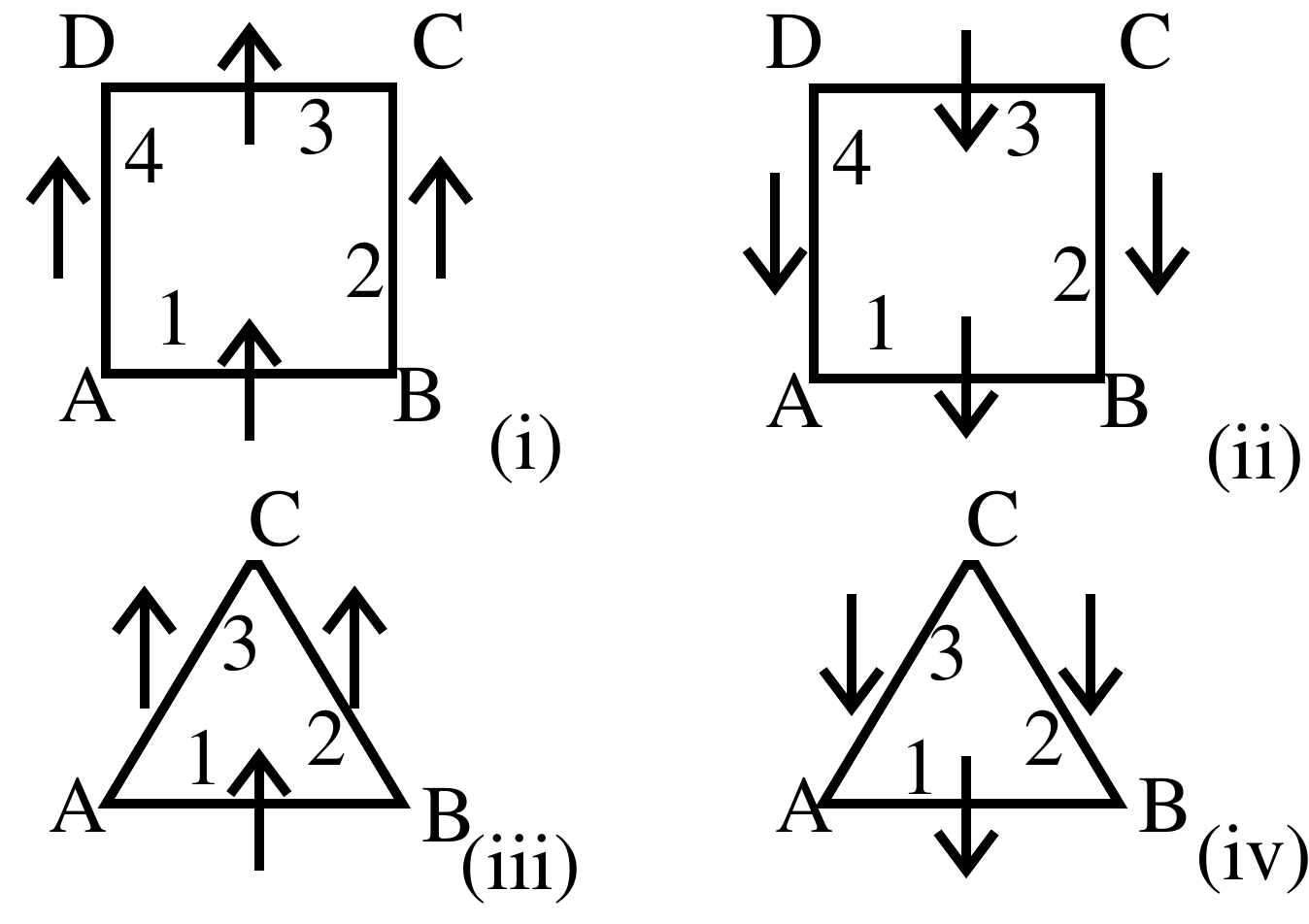}
      \caption{Basis states of the $\mathbb{Z}_2$ gauge theory in the $\SX{}$
      basis for both the square plaquette (upper row) and the triangular 
      plaquette (lower row). The configurations (i) and (ii) satisfy the 
      Gauss law $V_r = 1$ at all sites for the square, and the ones (iii) 
      and (iv) satisfy $V_r = 1$ at all sites for the triangular plaquette.}
      \label{fig:GS1}
  \end{figure}

 For the single-plaquette system shown in Figure \ref{fig:GS1} (top row) with four
 links in all and two links touching each vertex (labelled as A,B,C, and D), we start
 by explicitly writing the Hamiltonian and the Gauss law:
  \begin{equation}
  \begin{split}
      H & = -g~\SZ1 \SZ2 \SZ3 \SZ4 \, , \\
      V_{A} & = \Sx1 \Sx4;~V_{B}  = \Sx1 \Sx2;~ 
      V_{C} = \Sx2 \Sx3;~ V_{D}  = \Sx3 \Sx4.
  \end{split}
  \end{equation}
  For a single plaquette, 16 states are possible in total, which comprise the full 
  Hilbert space. We construct the Hamiltonian in each of the sectors characterized by 
  particular local values of the Gauss law. Since this a $\mathbb{Z}_2$ theory, the
  Gauss' Law can only take $\pm 1$ values. The two states illustrated in the top row of
  Figure \ref{fig:GS1} have $V_x |\psi\rangle =  1 |\psi\rangle$ at each site. 
  Similarly, it is possible to obtain two configurations which have 
  $V_x |\psi\rangle = -1 |\psi\rangle$ at each site. Furthermore, it is possible
  to place two positive and two negative $\mathbb{Z}_2$ charges, giving rise to 6 more 
  sectors. Each sector has two states which are related to each other by charge
  conjugation (global $\SX{} \leftrightarrow -\SX{}$ flip).

\begin{figure}
      \centering
      \includegraphics[scale=0.45]{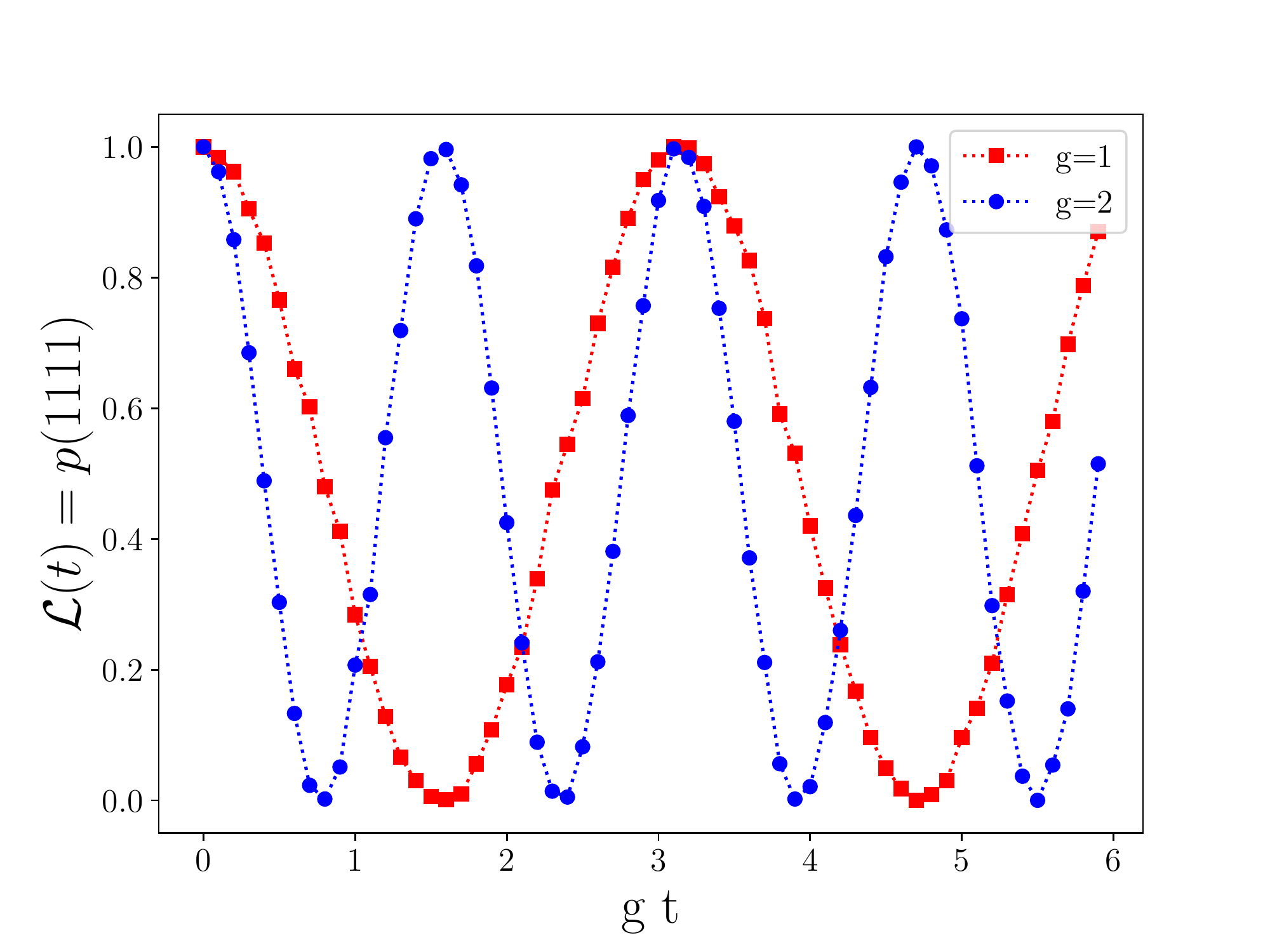}
      \caption{Oscillations of the Loschmidt probability ${\cal L}(t) = p(1111)$ for 
      the square $\mathbb{Z}_2$ plaquette on the ibmq\_qasm\_simulator, which is a general
      purpose simulator. The points are the points from the simulator, and the line is only
      to guide the eye. The system has a  two-dimensional gauge invariant Hilbert 
      space, and there is a two-state Rabi oscillation when started from the state 
      $|1111\rangle$ to the state $|0000\rangle$. An identical behavior is also 
      observed in the triangular $Z(2)$ plaquette. Increasing the coupling by a 
      factor of two is identical to speeding up the dynamics by a factor of two. 
    }
    \label{fig:oscillations_th}
  \end{figure}

  For our purposes, we consider the quench-dynamics within the sector 
  $(V_A, V_B, V_C, V_D) = (+,+,+,+)$. The Hamiltonian is two-dimensional in this sector with
  the eigenstates 
  \begin{equation}
  \begin{split}
      |\Psi_1\rangle & = (|1111\rangle + |0000\rangle)/\sqrt{2}, \\
      |\Psi_2\rangle & = (|1111\rangle - |0000\rangle)/\sqrt{2}. \\
  \end{split}
  \end{equation}
  Here the notation $|0000\rangle$ denotes all spins aligned in the $+1$ 
  direction of the $\SX{}$ (computational) basis, and $|1111\rangle$ denoting all spins 
  aligned in the $-1$ direction. Similarly, for the $(-,-,-,-)$ sector, we get,
  \begin{equation}
  \begin{split}
      |\Psi_3\rangle & = (|1010\rangle + |0101\rangle)/\sqrt{2}, \\
      |\Psi_4\rangle & = (|1010\rangle - |0101\rangle)/\sqrt{2}. \\
  \end{split}
  \end{equation}
  Again, the $0$'s and $1$'s denote spins aligned in the $+1$ and $-1$ 
 directions of the $\SX{}$ basis, respectively. The real-time evolution starting from 
 an initial state $|1111\rangle$ is therefore a two-state Rabi oscillation. 
 A useful  quantity to measure is the return or the Loschmidt probability, 
 defined as the projection of the  time-evolved initial state on to 
 the initial state:  
 \begin{equation}
 {\cal L}(t) = |{\cal G}(t)|^2;~~{\cal G}(t) = \langle \psi_0 | e^{-i H t} | \psi_0 \rangle.
 \end{equation}
 In Figure~\ref{fig:oscillations_th}, we show the return or the Loschmidt 
 probability, which is an indicator for the so-called dynamical quantum phase
 transitions \cite{Heyl_2019}. As shown in the figure, increasing the frequency
 is equivalent to speeding up the dynamics by the same factor. 

 It is also possible to consider the $\mathbb{Z}_2$ gauge theory on different lattices, 
  such as the triangular, hexagonal, or the checkerboard lattice. Here we will also
 consider the example of a triangular lattice. Again, considering a single plaquette 
 as illustrated in Figure \ref{fig:GS1} (below), there are three links in a plaquette, 
 and each vertex contains two links where the Gauss law can be imposed. In this case,
 labelling the three vertices as $A,B,$ and $C$; and the three links as $1,2,3$, the
 Hamiltonian and the Gauss law are:
\begin{equation}
\begin{split}
    H &= -g~ \SZ1 \SZ2 \SZ3, \\
    V_A &= \Sx1 \Sx2;~V_B=\Sx2 \Sx3;~V_C=\Sx3 \Sx1 \, .
\end{split}
\end{equation}
 The analysis of the triangular plaquette is also similar to the square plaquette,
 leading to two quantum states in each Gauss law sector (and four sectors total), 
 and thus the real-time evolution also displays a characteristic Rabi 
 oscillation similar to the one in the square plaquette. 

  In the following sections, we study both plaquette models on a quantum
 hardware, where decoherence will cause mixing among the different sectors. The extent
 of the mixing can help us to understand the (in-)efficiency of the quantum hardware,
 and which optimizations, error corrections or mitigations are likely to help.

 \subsection{The U(1) quantum link model}
\begin{figure}
    \centering
    \includegraphics[scale=0.28]{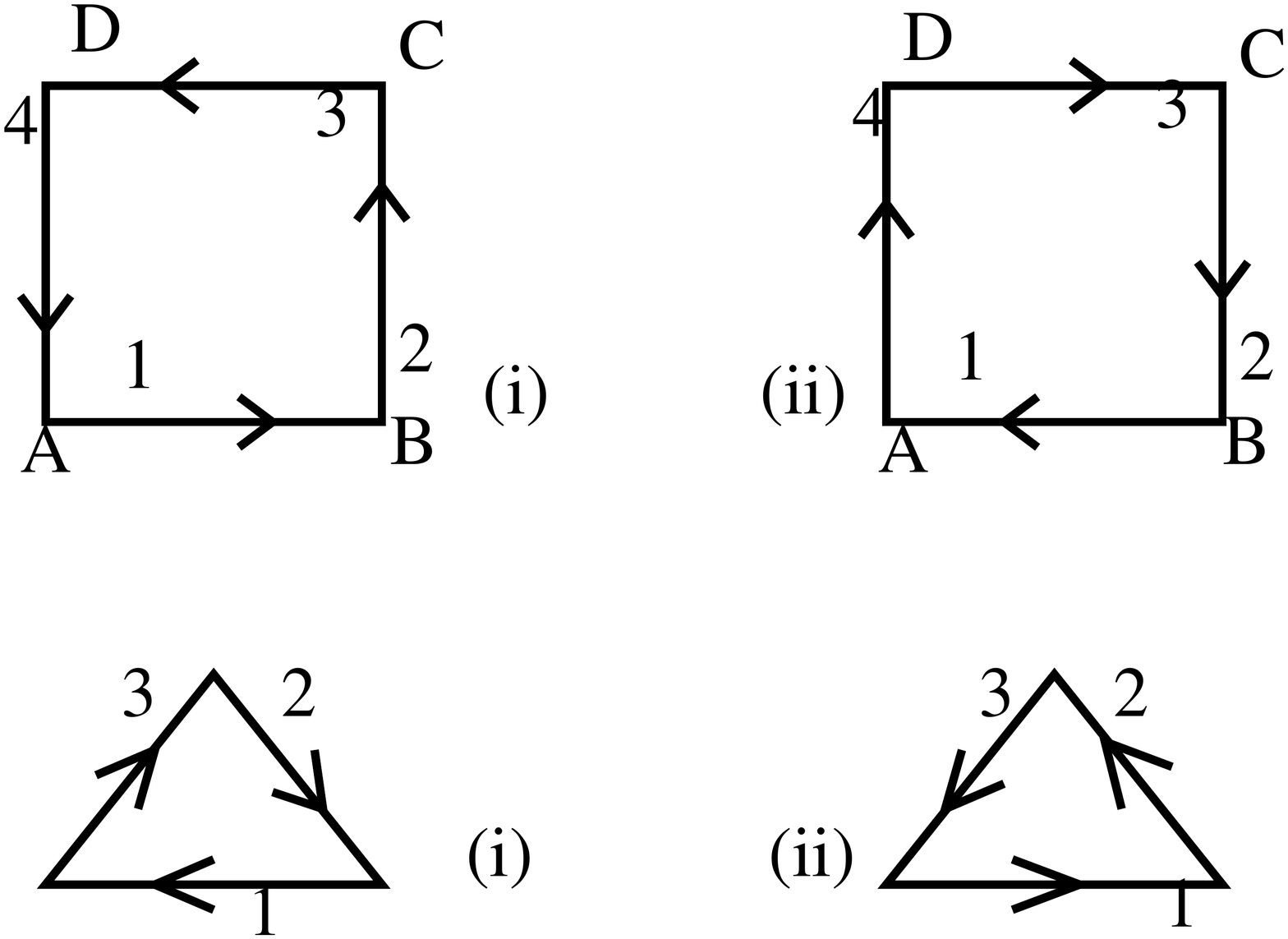}
    \caption{Sample basis states for the square (top) and triangular (bottom)
    plaquettes of the $U(1)$ QLM, where the spins are quantized in the $\Sz{}$
    basis. For the square lattice, the spins pointing up (down) indicated by arrows
    on the vertical links correspond to $E=+\frac{1}{2}(-\frac{1}{2})$. For the
    links along the x-axis (the horizontal links), the arrows pointing to the right
    (left) indicate spins quantized along $E = +\frac{1}{2} (-\frac{1}{2})$.  
    For the triangular plaquette, the arrows pointing in the clockwise (counter-clockwise) 
    direction indicate spins quantized along the $E = +\frac{1}{2} (-\frac{1}{2})$.
    Each of these examples of the basis states are in the $G_x = 0$ sector, which 
    can be seen physically from the fact that every point has one arrow coming in
    and another going out.}
    \label{fig:U1basis}
\end{figure}

  We next consider the case of the $U(1)$ lattice gauge theory, which has considerably
richer physics; and as a stepping stone to studying QED, has relevance to the fundamental 
physics of Nature. We will consider the theory on both the square and the triangular
lattice, as in the  case of the $\mathbb{Z}_2$ theory. The phase diagrams of both systems have
been studied in the literature \cite{Banerjee_2013, banerjee2021nematic}, as well as
aspects of dynamics and thermalization of the model on the square lattice 
\cite{Banerjee_2021} and its potential realization on analog and digital computers
\cite{Marcos_2014, Glaetzle_2015, Celi_2020}. Since we want to implement the models
using actual quantum hardware, we  will consider very small systems involving single
and double plaquettes, as shown in Figure \ref{fig:U1basis}.

  To implement a local $U(1)$ symmetry for the Hamiltonian in a simple way, we
need the spin raising and lowering operators, given by: 
$U_l = \SP{l} = \frac{1}{\sqrt{2}} (\Sx{l} + i \Sy{l})$ and 
$U^\dagger_l = \SM{l} = \frac{1}{\sqrt{2}}(\Sx{l} - i\Sy{l})$.
The operators $U_l$ (and $U_l^\dagger$) are canonically conjugate to the electric 
flux operator living on the same link, $E_l = \SZ{l}$, and obey the following 
commutation relations:
  \begin{equation}
      [E, U] = U;~~ [E, U^\dagger] = - U^\dagger;~~[U , U^\dagger] = 2 E \, .
  \end{equation}
Operators residing on different links always commute. With these operators, we 
can now define the lattice $U(1)$ Gauss law:
  \begin{equation}
    G_x = \sum_{\mu} \left( E_{x,\mu} - E_{x-\mu,\mu} \right).
    \label{eq:gl}
  \end{equation}
Note that $\mu$ denotes the lattice unit vectors, and thus for the square lattice
$\mu=1,2$, while for the triangular lattice $\mu=1,2,3$. This operator $G_x$ generates
the gauge transformations, which can be expressed as $V = \prod_x \exp \left(- i \alpha_x G_x \right)$, 
where $\alpha_x$ is the (local) parameter associated
with the local unitary transformation. This operator commutes with the plaquette
Hamiltonian defined on the entire lattice. For the square lattice, the local Hamiltonian
involves four links around a plaquette, and the model has the form
\begin{equation}
\begin{split}
 H_\Box   & = -g \sum_{\Box} \left( U_{\Box} + U^\dagger_{\Box} \right), \\
 U_{\Box} & = \SP{r,\mu} \SP{r+\mu,\nu} \SM{r+\nu,\mu} \SM{r,\nu},
\end{split}
\end{equation}
where $\mu,\nu$ are the lattice axes and $r$ is the bottom left corner of a square plaquette.
For the triangular lattice, the 3-link plaquette Hamiltonian has the form:
\begin{equation}
\begin{split}
H_\triangle &= -g \sum_{\triangle} \left( U_\triangle + U^\dagger_{\triangle} \right), \\
U_{\triangle} &= \SP{xy} \SP{yz} \SP{zx} ,
\end{split}
\end{equation}
where the points $x,y,z$ are the vertices of a triangle. Mathematically, the commutation
relation $[G_x, H] = 0$ ensures that the Hamiltonian is invariant under local unitary
transformations $H = V H V^{\dagger}$, resulting in a highly constrained system.

From these equations, the single-plaquette case can be obtained by only keeping the
links that exist in the triangle or the square geometry, and gives rise to:
  \begin{equation}
  \begin{split}
     H_\Box & = -g( \SP1 \SP2 \SM3 \SM4 + \SM1 \SM2 \SP3 \SP4 ); \\
     H_\triangle & = -g ( \SP1 \SP2 \SP3 + \SM1 \SM2 \SM3); \\
  \end{split}
  \label{eq:splaq}
  \end{equation}
  with Gauss law operators given by
  \begin{equation}
  \begin{aligned}
      G_A & = \SZ{4} +\SZ{1},\qquad G_B = \SZ{2} - \SZ{1}\\
      G_C & = -\SZ{2} - \SZ{3},\qquad G_D = \SZ{3} - \SZ{4},
      \end{aligned}
  \end{equation}
  for the square plaquette, and 
   \begin{equation}
      G_A  = \SZ{1} -\SZ{3},\;\; G_B = \SZ{2} - \SZ{1},\;\; G_C = \SZ{3} - \SZ{2}
  \end{equation}
  for the triangular plaquette, where the link subscripts correspond to the labels in 
  Figure \ref{fig:U1basis}. Note that the conventions for the signs of the electric flux 
  are given in the caption of the figure.
  
 For our purposes, it is useful to further simplify Equation (\ref{eq:splaq}) and 
 express the Hamiltonian in terms of the Pauli matrices, which will allow us to construct 
 the quantum circuits using the circuit identities introduced in the next section.
 For the square plaquette we obtain:
\begin{equation}
\begin{split}
 H_\Box  & = -\frac{g}{2} \left[ \Sx1 \Sx2 \Sx3 \Sx4 + \Sy1 \Sy2 \Sy3 \Sy4 - \Sx1 \Sx2 \Sy3 \Sy4 \right. \\ 
 &\qquad\qquad\left. - \Sy1 \Sy2 \Sx3 \Sx4  +  \Sy1 \Sx2 \Sy3 \Sx4 + \Sy1 \Sx2 \Sx3 \Sy4 \right. \\
 &\qquad\qquad\left. + \Sx1 \Sy2 \Sy3 \Sx4 + \Sx1 \Sy2 \Sx3 \Sy4 \right].
\end{split}
\label{explicitu1}
\end{equation}
  Thus there are eight terms for a single plaquette when expressed with the Pauli
  matrices. For the triangular plaquette, we have four independent plaquette terms which
  have to be implemented in a quantum circuit:
 \begin{equation}
 \begin{aligned}
  H_\triangle &= -g/\sqrt{2} \left[ \Sx1 \Sx2 \Sx3 - \Sy1 \Sy2 \Sx3\right. \\ 
  &\left. \qquad\qquad\quad- \Sy1 \Sx2 \Sy3 - \Sx1 \Sy2 \Sy3 \right].
 %\end{split}
 \end{aligned}
  \end{equation}

  The solution of the single-plaquette problem is straightforward: for the $U(1)$ system as defined here, 
 it is more natural to consider the system quantized in the $\Sz{}$-basis (instead of the $\Sx{}$-basis 
 used in the $\mathbb{Z}_2$ case), such that the spin-up and the spin-down can be denoted by arrows 
 pointing in and pointing out respectively from a given site. This can be interpreted physically as 
 the plaquette operators raising or lowering states by a unit of magnetic field (which is like a 
 clockwise or anti-clockwise arrangement of the electric fluxes around the plaquette). For the 
 triangular lattice this means that there are only $2^3=8$ basis states, and the square lattice has 
 $2^4=16$ such basis states. The Gauss law further selects only two basis states for each of the two 
 lattices. For the triangular lattice with $G_x=0$ everywhere as an example, we denote them as  $\ket{000}$ 
 and $\ket{111}$; while for the square lattice with $G_x=0$ we denote them as $\ket{0011}$ and $\ket{1100}$. 
 Note that $0$ denotes a spin-up and $1$ a spin-down in the $\Sz{}$ basis. The states are shown
 in Figure \ref{fig:U1basis}. The Hamiltonian for both cases is therefore a two-dimensional
 off-diagonal matrix. The two eigenstates are thus given by a symmetric and anti-symmetric linear 
 superposition of the two basis states. The real-time evolution -- with the Loschmidt probability 
 oscillating between the two basis states -- is qualitatively the same as that given in Figure 
 \ref{fig:oscillations_th}, the period simply differs as a function of $g$.

\subsection{Two-plaquette system}
\begin{figure}
    \centering
    \includegraphics[scale=0.5]{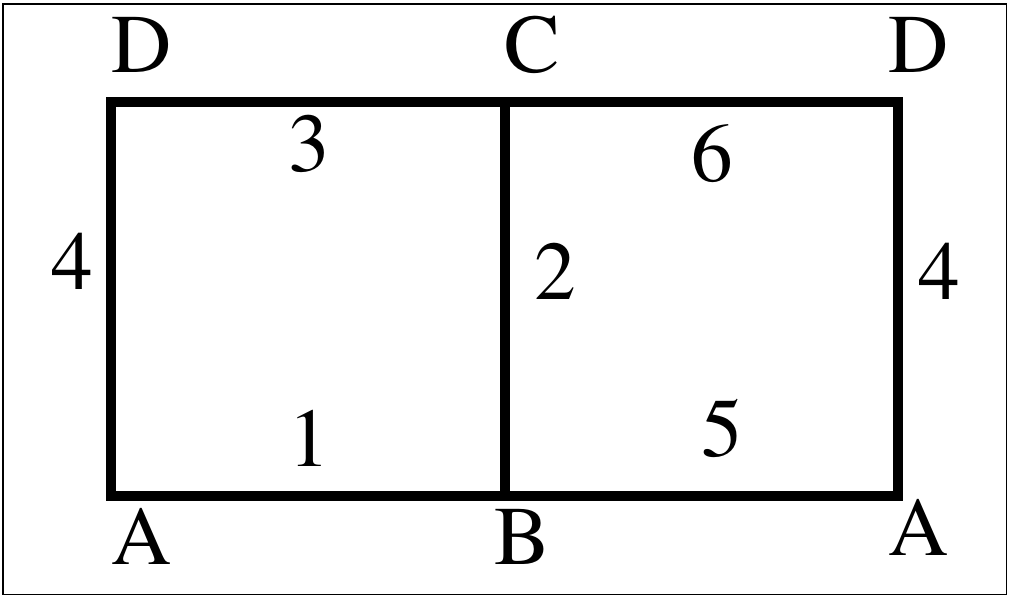}
    \caption{The set-up for two plaquettes which have periodic boundary conditions
    in the longer direction. The links are marked with numerals, while the sites
    are marked with letters. }
    \label{fig:2plaq}
\end{figure}

  As one more test of the quantum hardware, we consider a two-plaquette system on a
  square lattice with periodic boundary conditions for the $\mathbb{Z}_2$ gauge theory. The
  geometry of the system is shown in Figure \ref{fig:2plaq}.
 For clarity, let us explicitly write the Hamiltonian and the Gauss law for this case:
\begin{equation}
\begin{split}
     H & = -g \SZ1 \SZ2 \SZ3 \SZ4 - g \SZ5 \SZ4 \SZ6 \SZ2, \\
    G_{A} &= \Sx1 \Sx4 \Sx5; ~~~~G_{B} = \Sx5 \Sx2 \Sx1; \\
    G_{C} &= \Sx6 \Sx2 \Sx3; ~~~~G_{D} = \Sx3 \Sx4 \Sx6,
 \end{split}
 \label{eq:Z2_2plaquette}
 \end{equation}
 following the labeling in Figure \ref{fig:2plaq}. Because
 the $\sigma_{\rm N}^z$ commute with each other, the time evolution given by this
 Hamiltonian can be decomposed as the evolution given by the product of the time
 evolution given by each of the two terms for $H$ in Equation \eqref{eq:Z2_2plaquette}.
 This decomposition is exact and not subject to any Trotter errors. For each term, we can
 use the strategy to be described in the next section: introduce an ancillary qubit
 which couples to the rest of qubits in the plaquette, and perform dynamics with 
 the help of the ancillary qubit. Further, the structure of the Gauss law implies
 that we can impose the constraint $G_{x} = 1$ for all the sites. Without the
 constraint, there are $2^6 = 64$ states. The Gauss law constraint will then reduce
 this number. For example, imposing $G_{A} = 1$ affects the spins on the links 1, 4,
 and 5. Only those configurations are allowed where either all three have $+1$ in 
 the $\Sx{}$ basis, or exactly two of the spins 1,2, and 5 have $-1$ in the $\Sx{}$
 basis, and the third spin is $+1$. 
 
\begin{figure}
\centering
\includegraphics[scale=0.45]{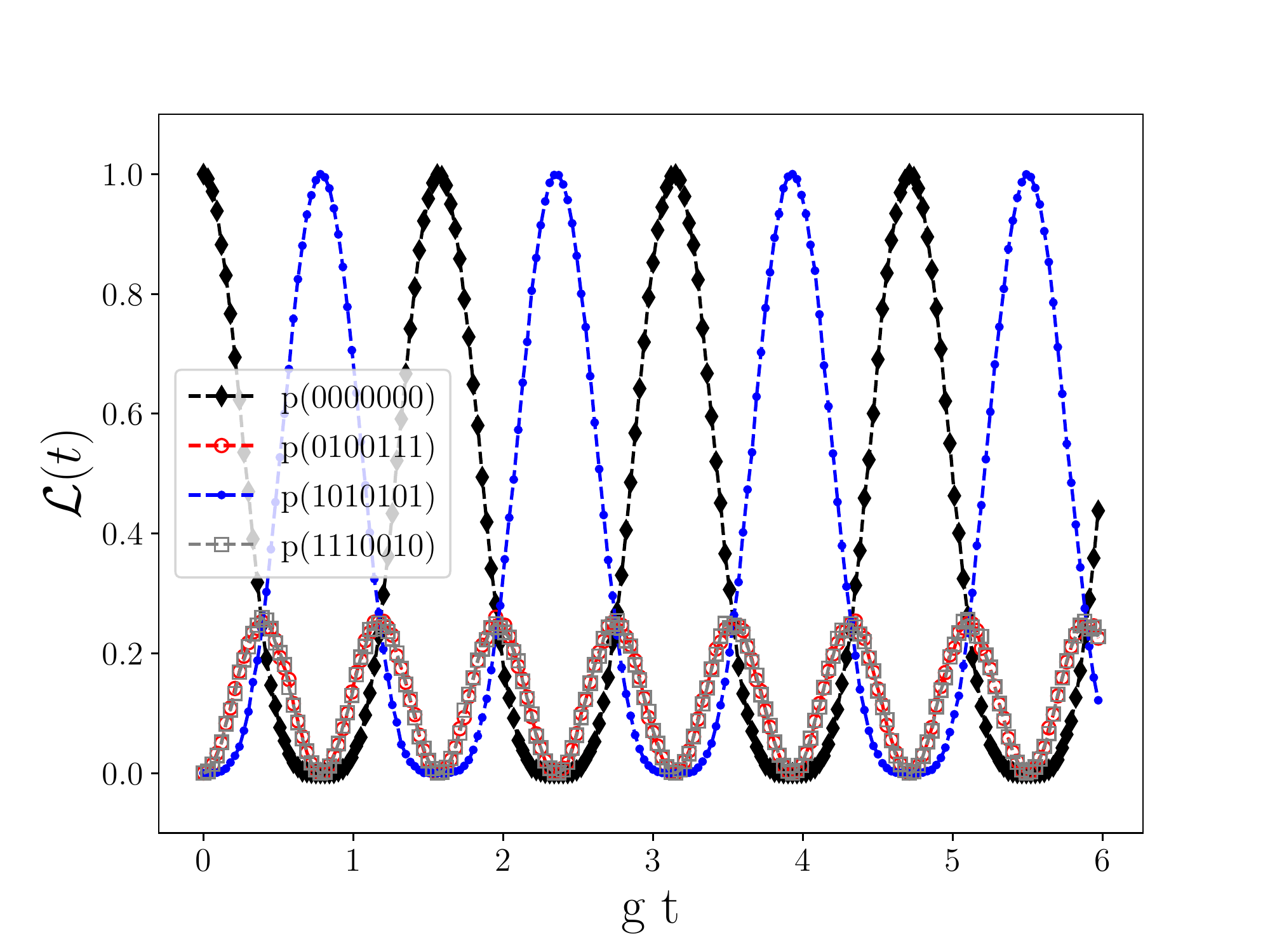}
\caption{Quench dynamics of the two-plaquette simulation from state 1 into the states
2,3, and 4, given by the ibmq-qasm-simulator. The Loschmidt probability oscillates 
between 0 and 1 for the states 1 and 3, while it oscillates between 0 and 0.25 for 
the states 2 and 4. Moreover, the probability oscillations between state 1 and 3 are 
exactly out of phase, as in the two-state systems considered previously, but it has 
equal projections into states 2 and 4. As before, the points are the ones from the
simulator, and the dashed line only guides the eye.}
\label{fig:2plaqSimulator}
\end{figure}
 While the solution of the two plaquette system is worked out in Appendix 
 \ref{ap:twoPlaquette}, we summarize the relevant points for the simulation of quench 
 dynamics of this system. The two plaquette system in the sector $G_{x} = 1$ for all 
 $x$ has 8 basis states. These 8 states can be further divided into two sectors using 
 the global winding number symmetry, which cuts the plaquettes horizontally and vertically 
 respectively. 
 
 For a general rectangular system with sizes $L_{\rm x} \times L_{\rm y}$, we can define 
 a global winding number $W_n$ for each of the spatial directions. If we draw a line at 
 a fixed $x=x_0$ ($y=y_0$) along the $y$($x$)-direction, then this line cuts all horizontal 
 (vertical) links (i.e. those pointing in the $x$($y$)-direction). Denoting the set of spins
 on the line as $\{ \sigma_m \}$, our winding number operator is given by
 \begin{align}
     W_n = \prod_{m} \sigma^x_{m},
 \end{align}
 where $n=y$ if $m=x$ and vice-versa. For our case, the expressions for the operators are 
  \begin{align}
      W_x & = \sigma^x_4 \sigma^x_2;~~~W_y(13)  = \sigma^x_1 \sigma^x_3;~~~W_y(56)  = \sigma^x_5 \sigma^x_6. 
      \label{eq:wind}
  \end{align}
  The last two expressions for $W_y$ are actually the same, as can be seen by using
  the Gauss law for the sites. Thus, in a perfect implementation, only 4 basis states
  entangle with each other under a unitary evolution. In Figure
  \ref{fig:2plaqSimulator} we show the Loschmidt probability for starting in one of
  these states, and the oscillations into the other three states. This system thus
  provides a good playground
  for tuning quantum hardware to reproduce these involved oscillations, as well as
  benchmarking to what extent local and global symmetries can be preserved in these
  circuits.
 
 For completeness, consider the $U(1)$ theory on two plaquettes, the entire Hamiltonian
 would have a total of 16 terms, which represented by the quantum gates are:
 \begin{equation}
 \begin{split}
 H &= -\frac{J}{2} \left[ \Sx1 \Sx2 \Sx3 \Sx4 + \Sy1 \Sy2 \Sy3 \Sy4 - \Sx1 \Sx2 \Sy3 \Sy4 \right. \\
  &- \Sy1 \Sy2 \Sx3 \Sx4 +  \Sy1 \Sx2 \Sy3 \Sx4 + \Sy1 \Sx2 \Sx3 \Sy4 + \Sx1 \Sy2 \Sy3 \Sx4 \\
  &+ \Sx1 \Sy2 \Sx3 \Sy4  + \Sx5 \Sx4 \Sx6 \Sx2 + \Sy5 \Sy4 \Sy6 \Sy2 - \Sx5 \Sx4 \Sy6 \Sy2 \\
  &- \Sy5 \Sy4 \Sx6 \Sx2   +  \Sy5 \Sx4 \Sy6 \Sx2 + \Sy5 \Sx4 \Sx6 \Sy2 \\ 
  &+ \left. \Sx5 \Sy4 \Sy6 \Sx2 + \Sx5 \Sy4 \Sx6 \Sy2 \right] .
 \end{split}     
 \end{equation}
 These terms do not all commute with each other, so trotterization would
 be necessary to simulate their real-time evolution. In this paper, we only
 consider the $\mathbb{Z}_2$ case which involves no Trotter steps.

  \section{Quantum Hardware and Circuits}
  \label{sec:circuits}
  In our plaquette model simulations, we make use of IBM Q hardware, which is based on
  superconducting (transmon) qubits. We discuss below a few details on how we work with
  this NISQ hardware, both in terms of selecting the platform for each experiment and
  in terms of circuit implementation.
  
  \subsection{Hardware Selection}
  Superconducting qubits have the advantage of being relatively fast at running
  experiments compared to trapped-ion qubits, but a disadvantage of relatively short
  decoherence times\cite{Linke3305}.
  
\begin{figure}[h]
      \centering
      \includegraphics[width=7cm]{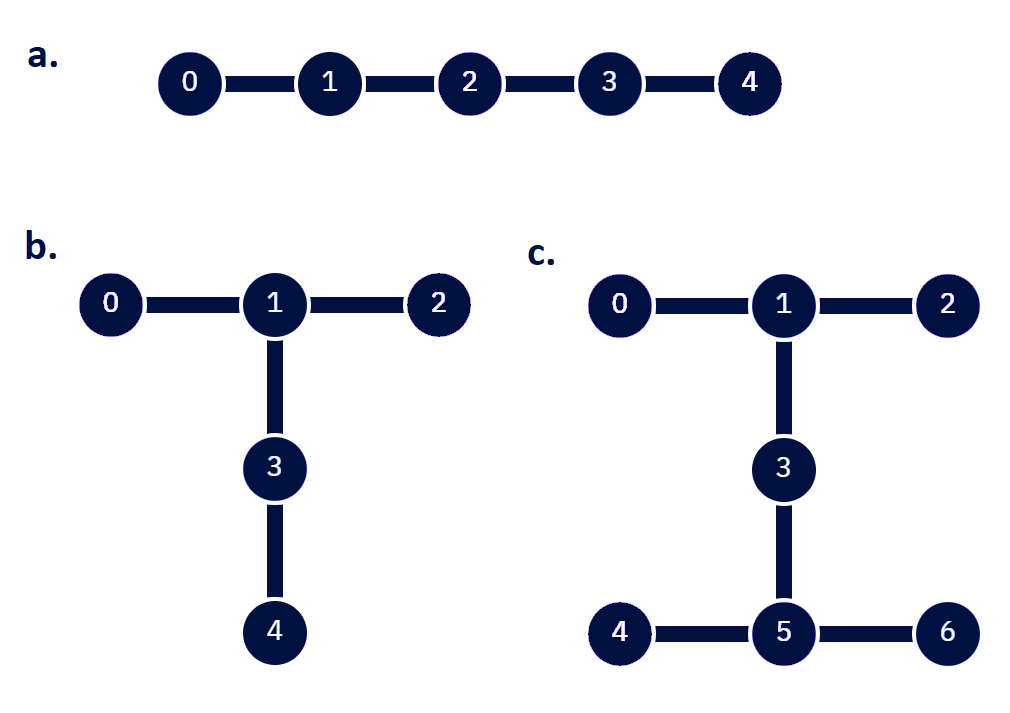}
      \caption{Three circuit topologies used for the simulations. Images taken from 
      IBM Quantum Experience.}
      \label{fig:topology}
  \end{figure}

  Because of this, the topology of the circuits is important, as it will make a
  difference for how many gates are necessary to realize a particular simulation.
  Figure \ref{fig:topology} shows three real-hardware topologies that are used in this
  paper. For each experiment, we may select hardware depending on optimal topology.
  
  Another important consideration for choosing hardware is the \textit{quantum
  volume} of the device, which is generally a measure of the most complex circuit 
  that can compute accurate quantities according to a
  particular threshold for a given device.
  IBM Q measures quantum volume using the following formula,
  \begin{equation}
      V_Q = 2^{{\rm min}(d,m)},
  \end{equation} where $d$ is the depth of the circuit (measured according to two-qubit
  gates), and $m$ is the number of qubits, so that ${\rm min}(d,m)$ tells us the
  largest square circuit possible that still meets the set accuracy threshold
  \cite{PhysRevA.100.032328}. The IBM Q devices each have a $V_Q$ measured and so in
  our experiments we favor using those with the higher $V_Q$ values. Specifically, 
  the devices used to obtain our results include IBM Q Valencia and IBM Q Quito, 
  which each have $V_Q=16$, as well as IBM Q Bogota, IBM Q Santiago, and IBM Lagos, 
  which each have $V_Q=32$.
  
  \subsection{Circuit Implementation and Scaling}
   The real-time simulation of plaquette dynamics involves realizing Hamiltonians
   of several spins on a plaquette. A very simple case looks like
   \begin{equation}
       H_{\rm N} = -g \sigma^3_{xy} \sigma^3_{yz} \sigma^3_{zw} \sigma^3_{wx},
   \label{eq:hex}
   \end{equation}
   with ${\rm N}=4$ and the sites ${x, y, z, w}$ are corners of a square
   plaquette. To realize a real-time evolution with the above Hamiltonian, we implement the
   following gate sequence \cite{2011NJPh13h5007M,Mezzacapo:2015bra}
   
   \begin{equation}
   \begin{aligned}
       U_{\rm S,A} (t) &= \exp \left[ i \frac{\pi}{4} \sigma^3_{\rm A} 
       \sum_{j=1}^N \sigma^3_j \right] \exp \left[ i g t \sigma^1_{\rm A}\right] \\
       &\qquad \qquad \times\exp \left[- i \frac{\pi}{4} \sigma^3_{\rm A} 
       \sum_{j=1}^N \sigma^3_j \right].
       \label{eq:qgate1}
       \end{aligned}
   \end{equation}
    A proof for Equation~\ref{eq:qgate1} is detailed in Appendix \ref{ap: circuitProof}.
    
 This identity has the nice property of being applicable to plaquettes that contain a 
 general number of spins, $N$, and in all cases allows for the time-evolution portion 
 to be done entirely on a single extra spin, which we label with the index $A$. This 
 spin is in addition to the $N$ spins that make up the plaquette, and is known as 
 \textit{ancillary}. In principle for a quantum circuit implementation (where we represent 
 each spin with a qubit), one would only need one ancillary qubit for the entire
 system, but due to topological issues it may be more efficient in terms of circuit
 depth to add more ancillary qubits in systems with more plaquettes. Still, with at
 most one ancillary qubit per plaquette, the number of qubits needed for simulation
 scales linearly with the number of links in the system.
    
  If all terms in the Hamiltonian commute, the number of gates needed is constant as a
 function of real time, but in the more generic case where the terms do not commute and
 so trotterization is necessary, the circuit depth scales linearly with time. In
 our examples below we focus only on cases where no trotterization is needed.

 \begin{figure}[h]
   \centering
   \includegraphics[width=8cm]{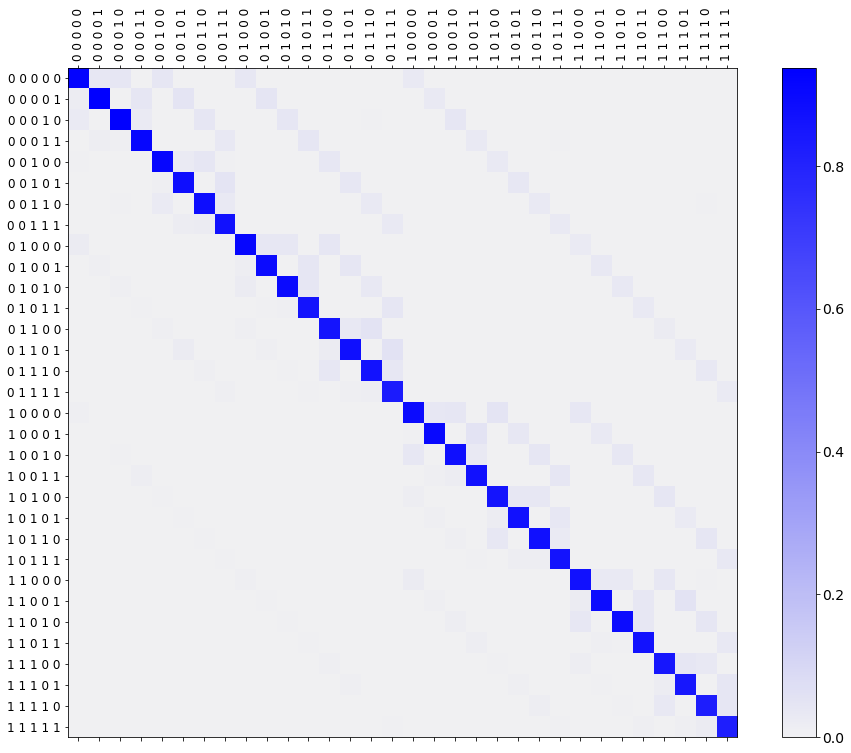}
   \caption{Response mitigation matrix computed by \texttt{ignis} for IBM Q Manila 
   (5 qubit system).}
   \label{fig:output}
 \end{figure}
 
 \section{Error Mitigation methodologies}
 \label{sec:errcorr}
 As mentioned earlier, one major practical obstacle to develop physical devices 
 to perform quantum computations is the significant inherent noise that affects NISQ quantum 
 devices. In theory, quantum error correction is possible by encoding the information 
 of the desired circuit into a highly-entangled state composed of a very large number 
 of physical qubits \cite{PhysRevA.52.R2493, PhysRevLett.77.793}. However, this large 
 number of qubits makes the hardware requirements too demanding to be implemented in 
 practice (although promising results point in the right direction \cite{Chen2021}). 
 An alternative is to take advantage of systematic and reproducible properties of the 
 hardware. These properties are exploited as part of the so-called error mitigation 
 schemes, which have proven to be successful in NISQ era devices 
 \cite{PhysRevX.7.021050, Kandala_2019,He_2020, larose2020mitiq, Giurgica_Tiron_2020, lowe2020unified, sopena2021simulating, funcke2020measurement, Nachman2020, Jattana2020, 2020QS&T5cLT01G}. 
 Among those, we consider two types, readout error mitigation and zero noise 
 extrapolation (ZNE); which aim to reduce noise coming from two different sources: 
 readout and gate operation decoherence. We emphasize that while here we use 
 these techniques on IBM Q hardware, they are in fact hardware-agnostic techniques -- 
 they rely only on the set of gates available and do not rely on the details of the 
 hardware such as the type of qubits used--and therefore can be used to improve results 
 on any universal quantum device.

 \subsection{Readout error mitigation}
 One important source of errors are the so called ``readout'' errors, which arise
 due to the comparable measurement and decoherence times 
 \cite{funcke2020measurement, Maciejewski2020, Jattana2020, Nachman2020}. This can 
 cause undesired
 state decays, affecting the state captured in the measurement. Assuming a 
 classical stochastic model for the noise affecting measurements, 
 the statement of the problem can be
 formulated by using the response matrix $P(m | t)$, which connects a 
 noisy measurement $m$ to the true/ideal measurement $t$ by the relation $m =
 P t$. Naively one can use the inverse of the response matrix to obtain $t =
 P^{-1} m$ and recover the true value of the measurement. The problem then consists in 
 performing a series of calibration experiments to measure $P$,
 and then use it to recover $t$ given $m$ in subsequent independent experiments.
 
 Packages such as qiskit-ignis \cite{Qiskit} are based on the response matrix formulation 
 of the readout error mitigation scheme, but (by default) do not try to compute $P^{-1}$ 
 directly by matrix inversion. 
 Instead, $t$ is recovered by finding the minimum of the least squares expression:
 \begin{equation}
     f(t) = \sum_{i=1}^{2^n} \left(m_i - \left(P \cdot t\right)_i \right)^2 \, ,
 \end{equation}
 where $n$ is the total number of qubits in the circuit. This methodology is 
 more robust than matrix inversion for general NISQ hardware \cite{Qiskit, Nachman2020}. 
 More involved methods
 combine the previous approach with gate inversion to further improved the error
 mitigation results \cite{Jattana2020}; while unfolding methods have also been proposed
 and tested in the literature \cite{Nachman2020}.
 
 In most cases, the ability to apply readout error mitigation is limited by the
 number of qubits ($n$) in the circuit, as the number of calibration experiments required to
 evaluate $P$ grows as $2^n$. Moreover, the calibration step of estimating $P$ is
 hardware dependent and needs to be performed immediately before running the
 experiments to guarantee temporal deviations in the particular hardware are accounted for.
 An real-hardware example of the response matrix obtained for
 a 5 qubit system (IBM Q Manila) using \texttt{ignis} is shown in \ref{fig:output}. 
 As expected, the diagonal entries
 have probability values close to 1, but there is still significant drift towards non-diagonal
 entries. As presented and discussed in Sec \ref{sec:results}, correcting for these small 
 deviations resulted in significant improvements in the final mitigated data.
 
Clearly, going beyond circuits with a small number of qubits would be prohibitively
expensive due to the number of experiments required to evaluate the response
matrix. Some proposals have considered the possibility of assuming close to
uncorrelated readout 
errors between the qubits, which would drastically reduce the number of experiments 
required \cite{Maciejewski2020}. 
Studying these potential improvements goes beyond the scope of this work.

 \begin{figure*}[!ht]
 \centering
 \includegraphics[width=12cm]{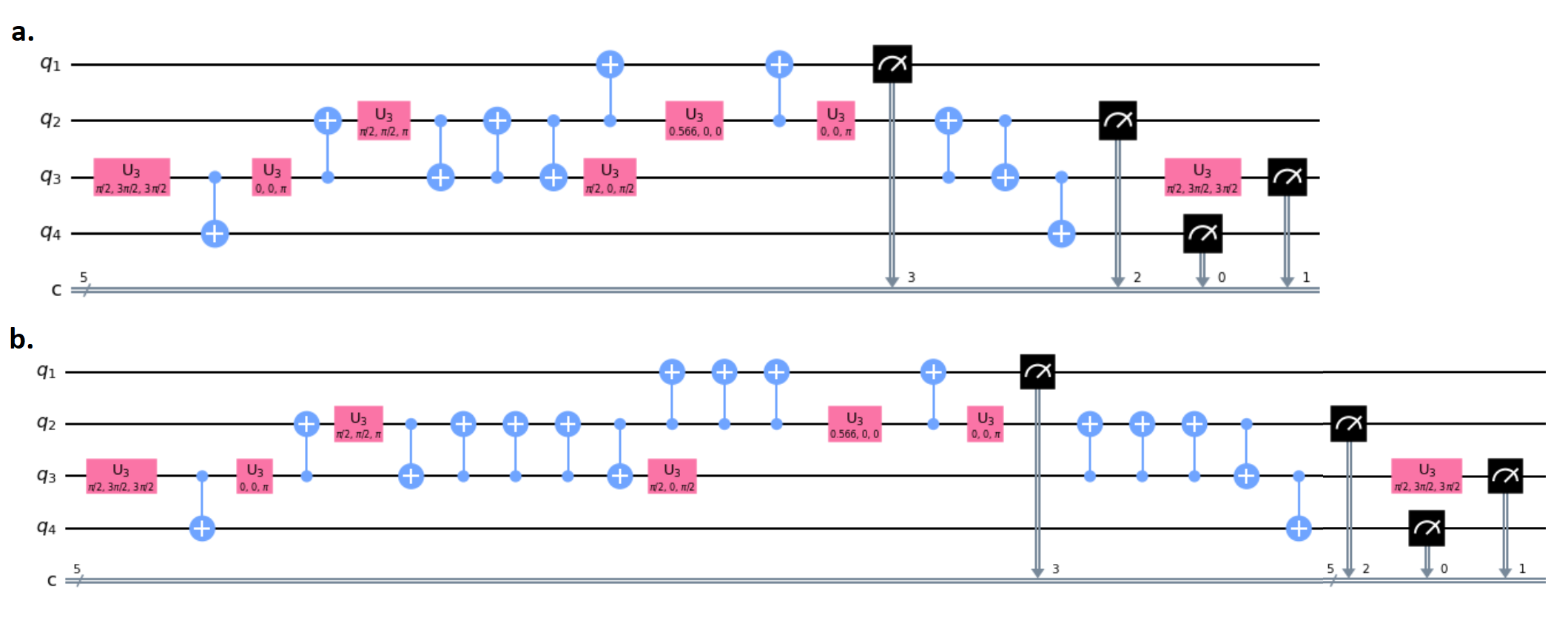}
 \caption{An example of using the Mitiq package for folding a circuit that gives the
 time evolution of the $Z(2)$ gauge theory on a triangular plaquette. Both circuits are
 equivalent, but the second one contains additional identity insertions of CNOT gates
 such that when measured using the CNOT circuit depth the second circuit is 1.6 
 times longer than the former.}
 \label{fig:folding}
\end{figure*}

 \subsection{Corrections against decoherence -- Mitiq}
The second source of error comes from the gate portion of the circuit before
measurements occur. Longer circuits will consist of more gates, and both the longer
runtimes and the gate implementation (transmon qubits in the case of the IBM Q devices)
will
cause additional errors to pile up. To mitigate this source of error we use a method
known as zero-noise extrapolation (ZNE), where we introduce additional noise in a
controlled way in order to empirically develop a noise model that we can extrapolate
to the zero noise case.

Implementations of ZNE include those that involve pulse control and run multiple 
experiments with pulses of different durations \cite{Kandala_2019}, and those that
involve \textit{folding}, which consists of insertions of additional gate identities 
to the circuit which would not change the results in an ideal simulation, but will 
make results on real hardware more noisy. This information on how the gates affect 
the noise level can then be used to develop a noise model and extrapolate back to 
an ``ideal'' result.

We used the folding option in this paper and specifically we used the \textit{Mitiq}
package to implement it \cite{larose2020mitiq}. As an example, Figure \ref{fig:folding}
shows two equivalent circuits, but the second circuit has three extra identity
insertions, each consisting of two identical CNOT gates in a row. Because the error
rates of the two-qubit CNOT gates are significantly higher than those for the single
qubit gates (roughly ten times different on IBM Q devices), we will assume perfect
fidelities for the single qubit gates and model all the error coming from the 
two-qubit gates (an option within \textit{Mitiq}). With this in mind, because circuit
$a$ in Figure \ref{fig:folding} has ten CNOT gates, and circuit $b$ has
sixteen CNOT gates, the scale factor of the circuit $b$ is 1.6 times that of
circuit $a$.

\begin{figure*}[b]
      \centering
      \includegraphics[width=6cm]{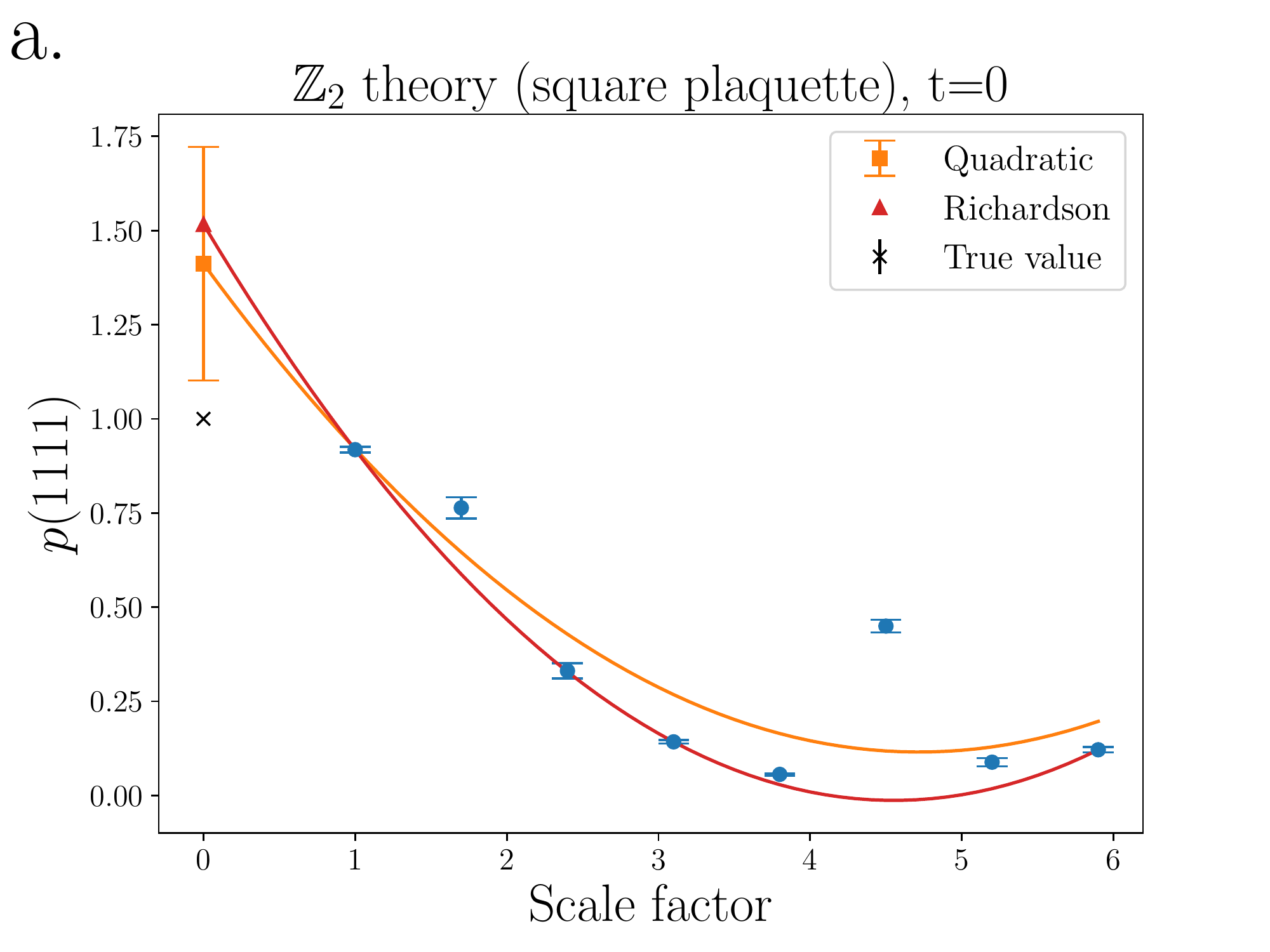}
      \includegraphics[width=6cm]{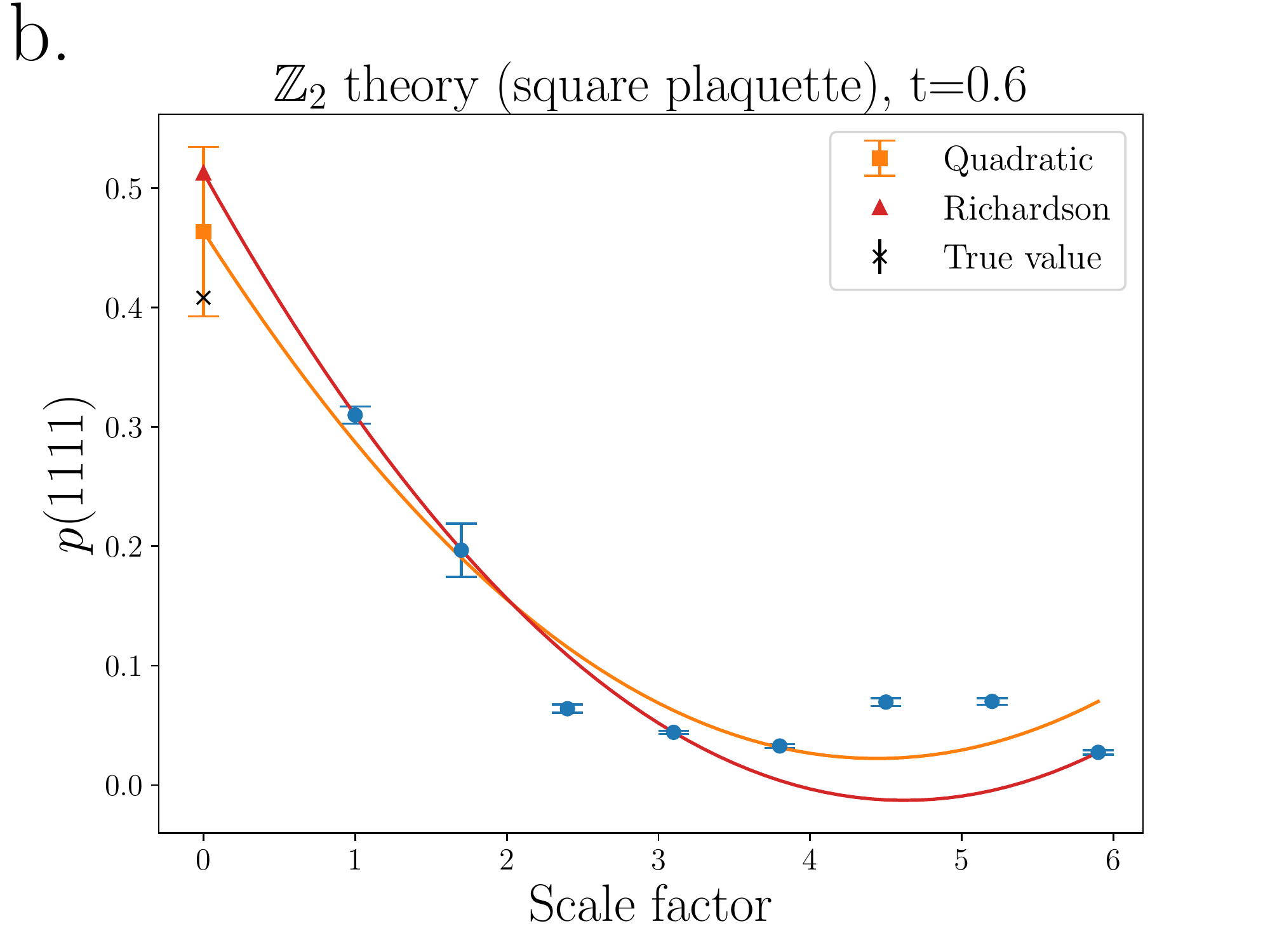}
      \includegraphics[width=6cm]{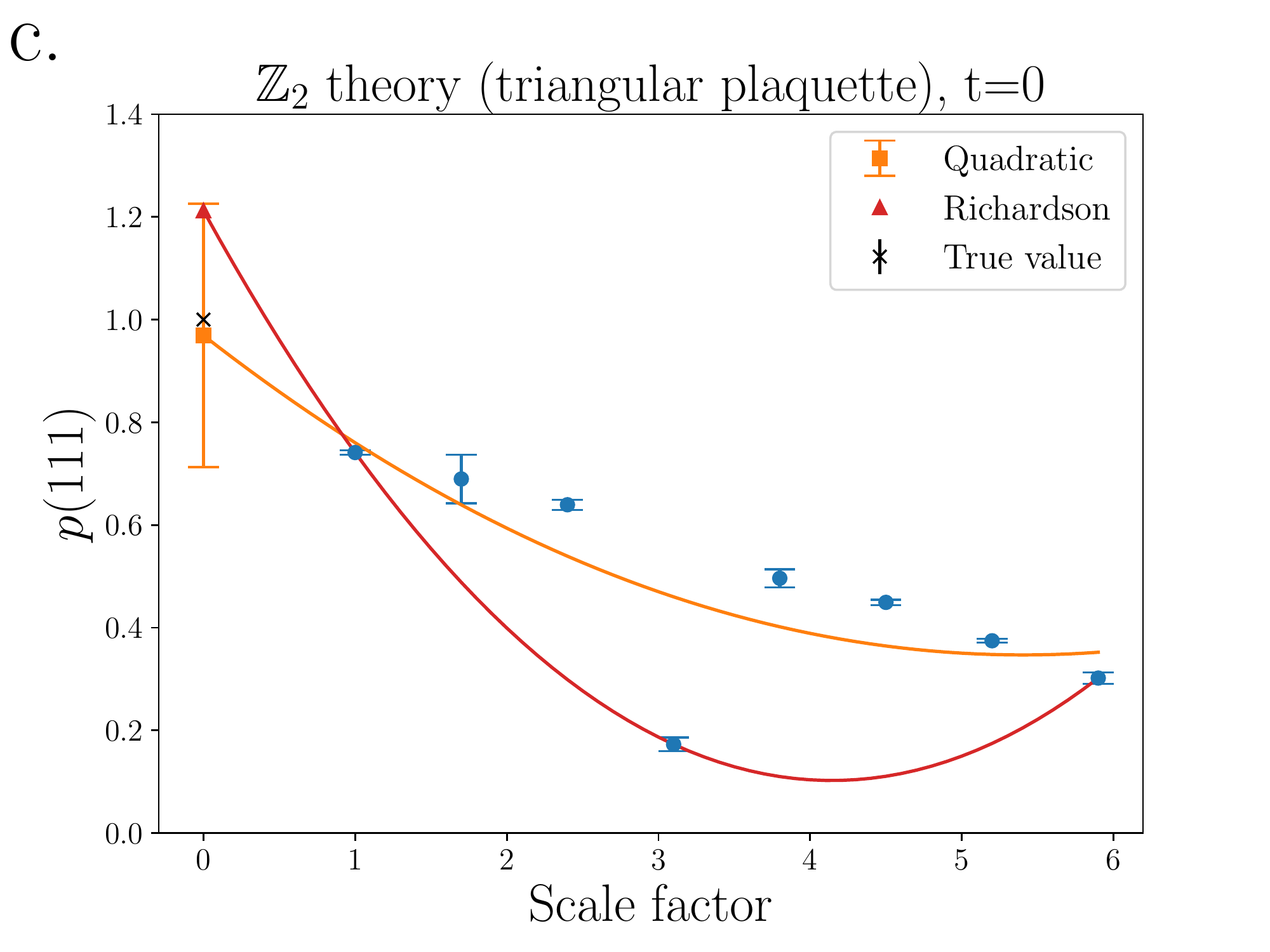}
      \includegraphics[width=6cm]{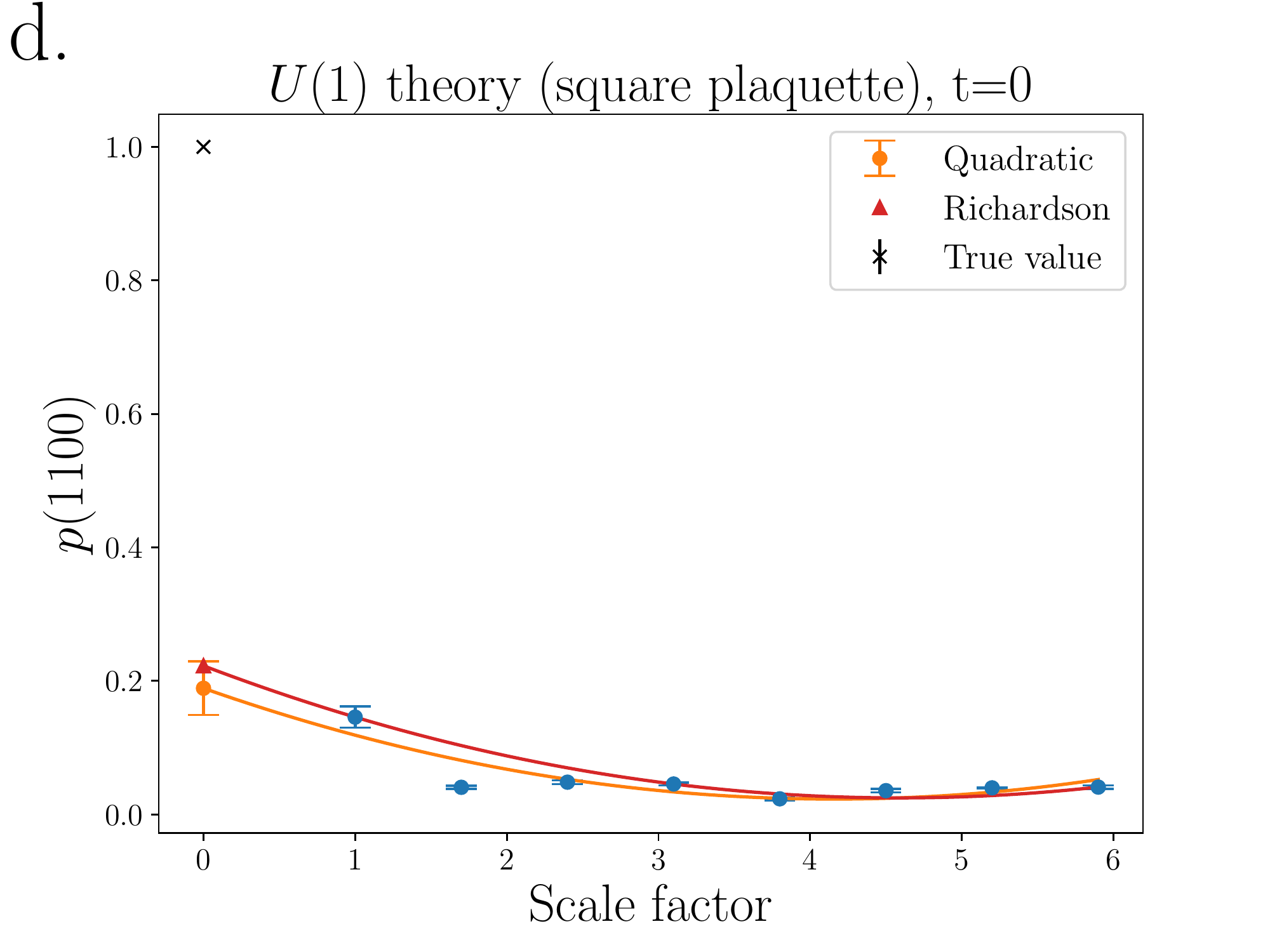}
      \caption{The plots in the top row show zero-noise extrapolation for the
      $\mathbb{Z}_2$ theory on a square plaquette (IBM Q Valencia hardware) at two
      times: $t=0$ ($a$) and $t=0.6$ ($b$). The bottom row shows zero-noise
      extrapolation for  the $\mathbb{Z}_2$ gauge theory on a triangular plaquette 
      (IBM Q Bogota) at $t=0$ ($c$) and a U(1) gauge theory on a square 
      plaquette (IBM Q Quito) at $t=0$ ($d$).}
      \label{fig:zne}
  \end{figure*}

 Figure \ref{fig:zne} shows real-hardware examples of different extrapolations for several
 circuits with the ideal result for each (determined using a simulator) marked at scale factor
 ``0''. The first row shows example extrapolations for $\mathbb{Z}_2$ model on the
 square plaquette, at two different times in the evolution. The bottom left image shows
 an extrapolation at $t=0$ for the $\mathbb{Z}_2$ theory on the triangular plaquette,
 and the bottom right image shows one at $t=0$ for the U(1) theory on the square
 plaquette. The two extrapolations shown are a quadratic fit and a Richardson
 extrapolation, explained in Kandala et. al.\cite{Kandala_2019} From this empirical
 data we decided to use the quadratic extrapolation for our data, as it appeared less
 susceptible to experimental outliers (such as those in the bottom left of Figure
 \ref{fig:zne}).

  It is interesting to note the presence of two regimes which display sensitivity
 to a change in the circuit depth. For larger scale factors which exceed the
 quantum volume of the system, the dependence on the scale factor becomes insensitive.
 At $t=0$, the measurements for increasing the circuit length decay only slowly
 until the scale factors exceed $3$ for the $\mathbb{Z}_2$ model, and about $6$ for
 the $U(1)$ model. For $t=0.6$ this decay is much faster for the $U(1)$ model than 
 the $\mathbb{Z}_2$ model. Typically the $U(1)$ circuit is significantly more entangled,
 and becomes more so when the extrapolation is attempted at finite $t$.

\section{Results}
\label{sec:results}
 This section gives our real-time evolution results for the Loschmidt probability, as well as
 observables $G_x$ and $W_y$ for plaquette simulations on NISQ hardware. In each
 simulation,  we take five measurements (8192 shots per measurement) at every point in
 time and at each of the eight different scale factors illustrated by Figure
 \ref{fig:zne}. This allows us to get error bars and perform ZNE at every time. 
 For each time, the different scale factor measurements were all taken within the same 
 calibration: see Appendix D for a note about the fluctuations of the measurements across 
 different calibrations of the IBM Q hardware. Each simulation consists of 20 points in 
 time total, leading to $5\times8\times20=800$ circuit measurements to produce the 
 error-mitigated plots for a theory on a particular plaquette.

\subsection{\texorpdfstring{$\mathbb{Z}_2$}{Z(2)} Theory on Single Plaquettes}
 \begin{figure*}
      \centering
      \includegraphics[width=5.75cm]{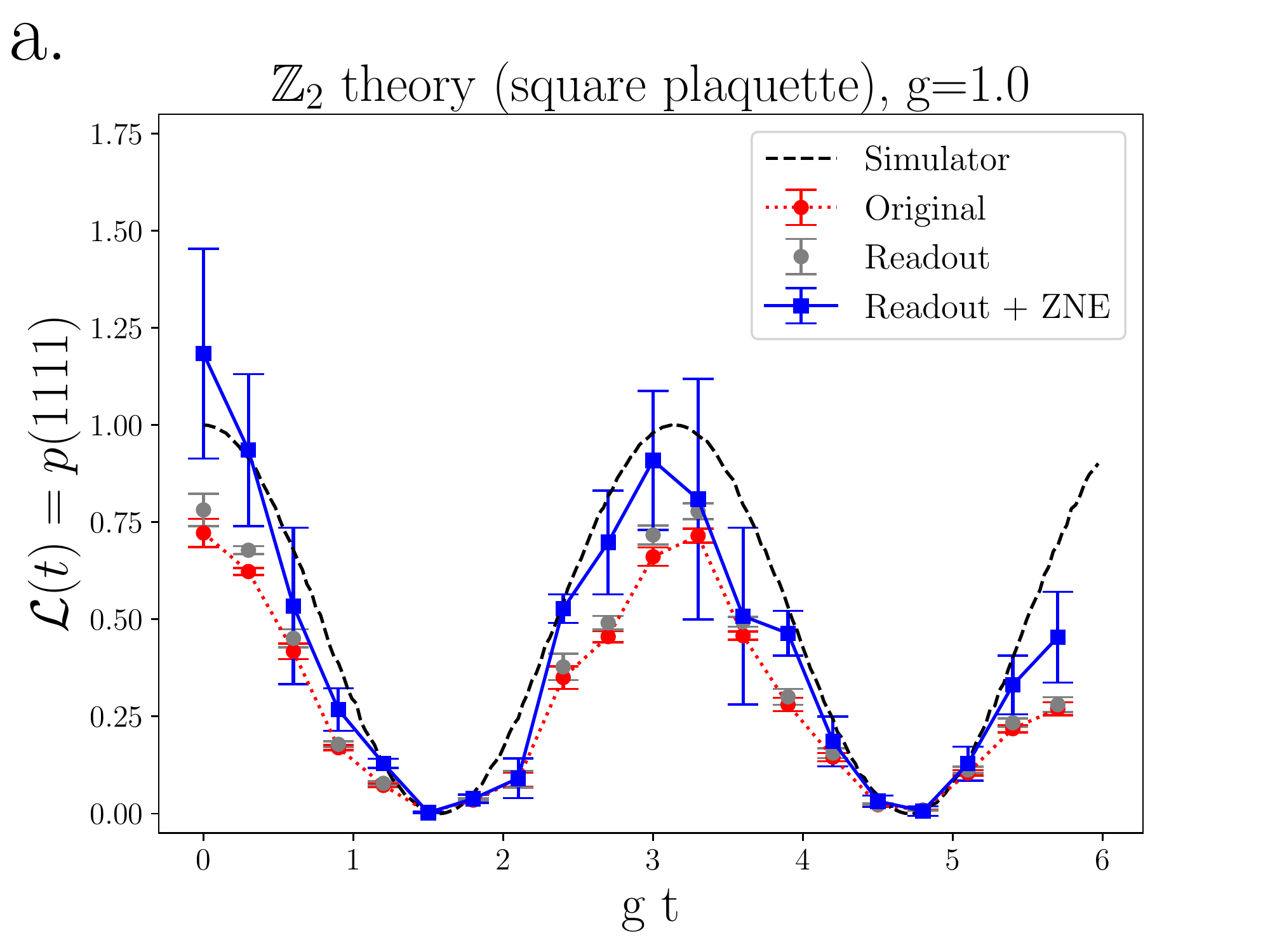}
      \includegraphics[width=5.75cm]{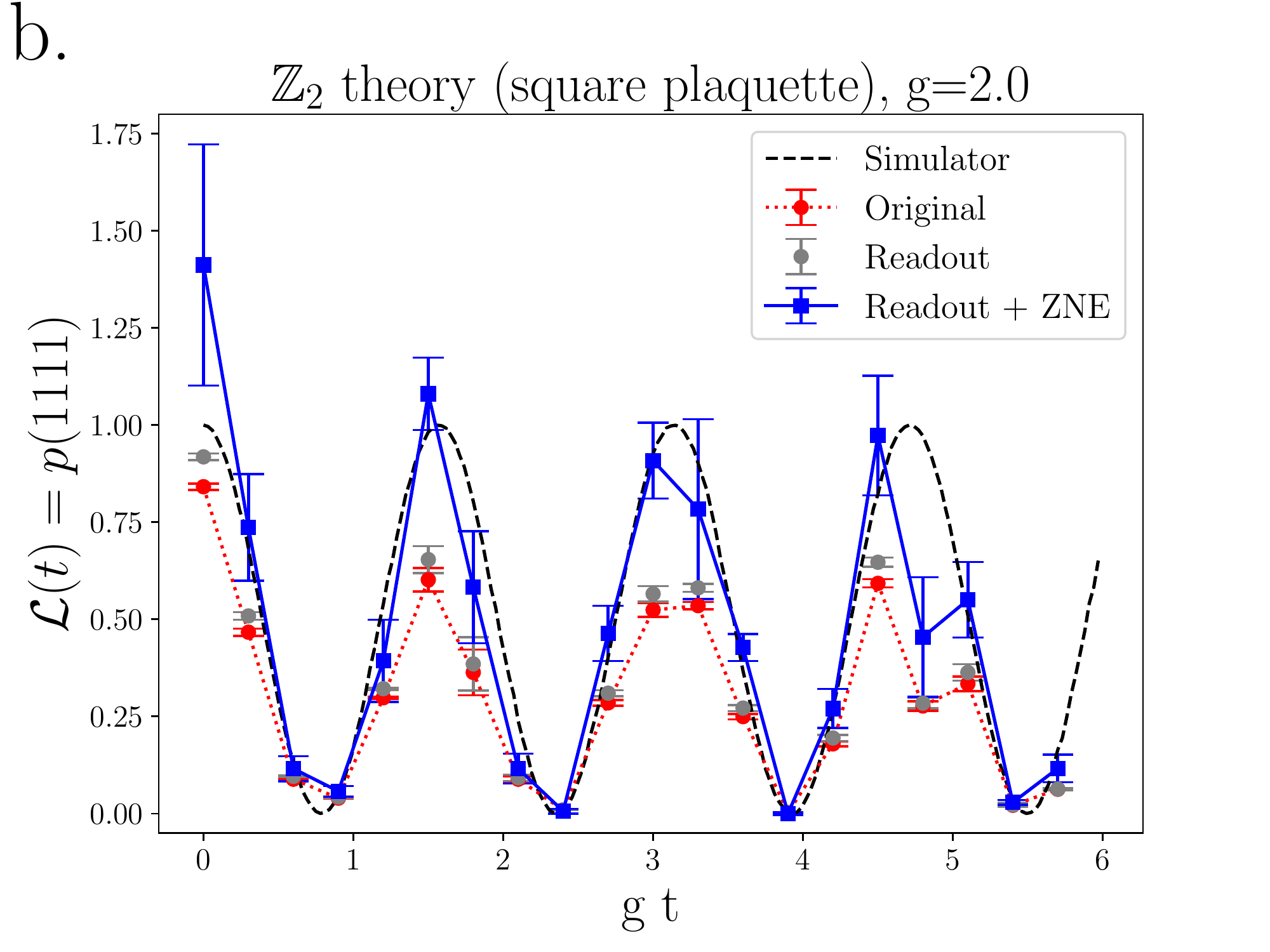}
      \\
      \includegraphics[width=5.75cm]{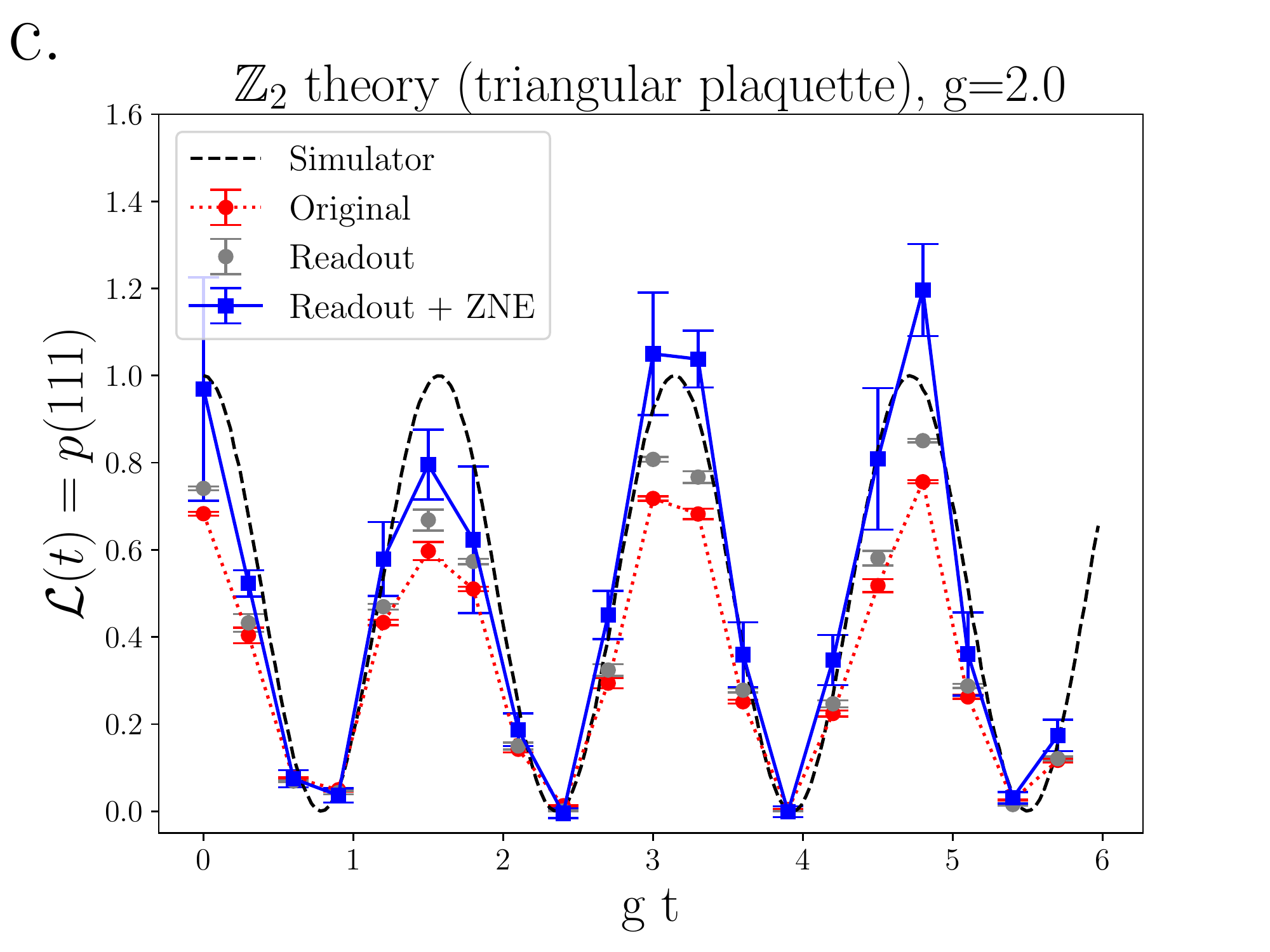}
      \includegraphics[width=5.75cm]{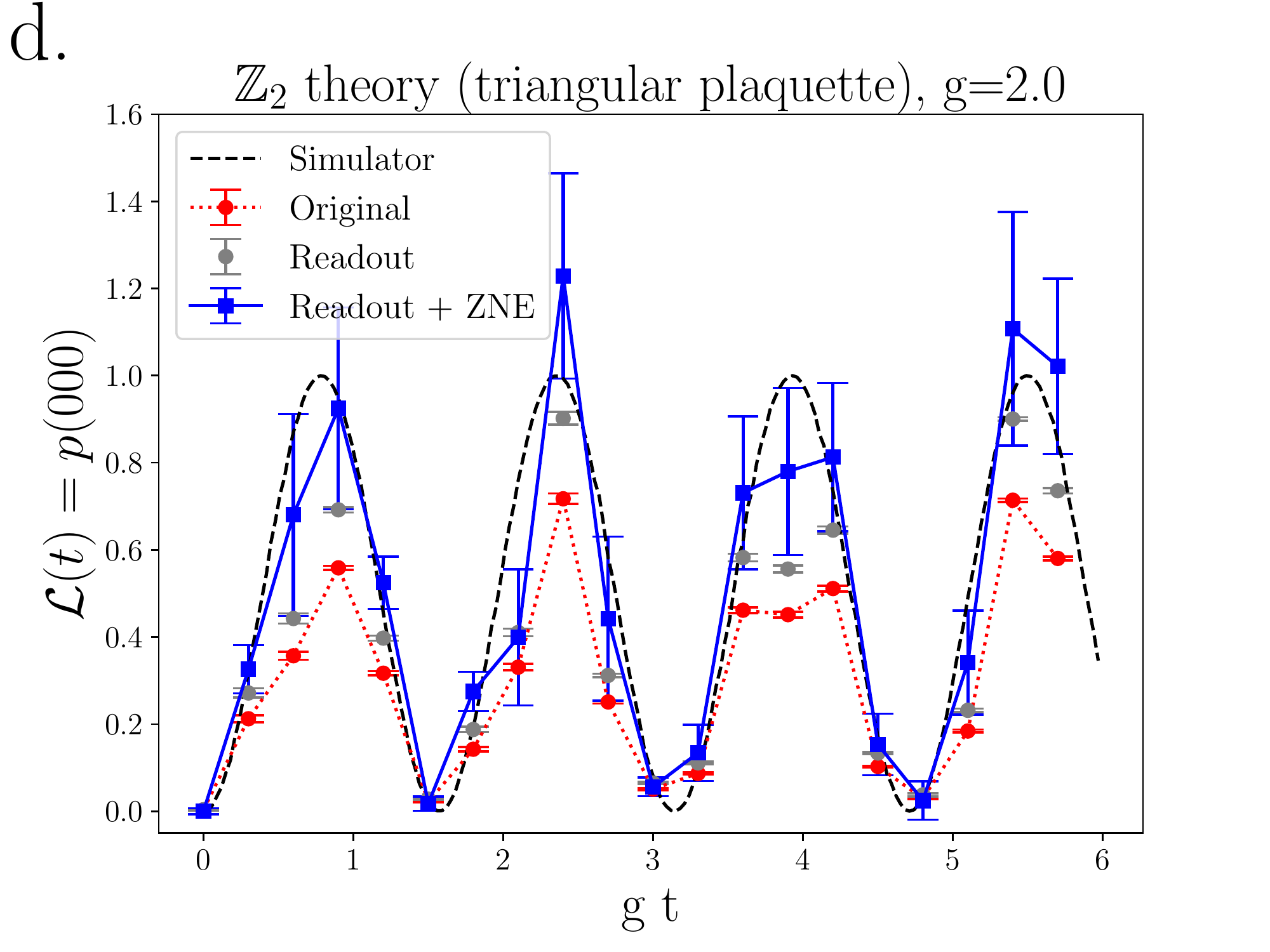}
      \includegraphics[width=5.75cm]{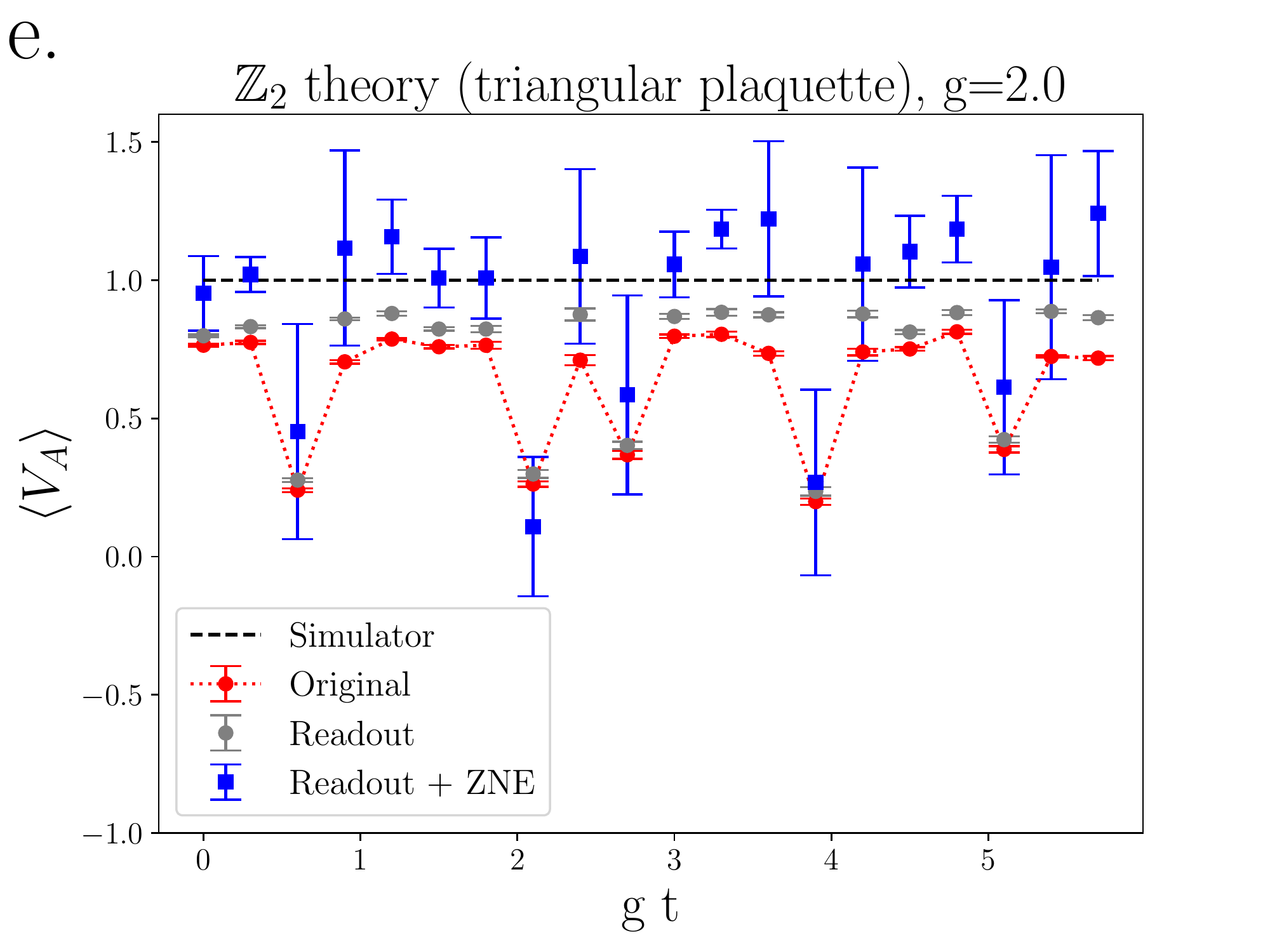}
      \caption{Real-time evolution of the $\mathbb{Z}_2$ theory on a single plaquette.
      Plots $a$ and $b$ in the first row show the Loschmidt probability data for a 
      square plaquette on IBM Q Valencia (with two couplings: g=1.0,2.0), then plots 
      $c$ and $d$ show the Loschmidt probability data for a triangular plaquette on IBM
      Q Bogota. Finally, plot $e$ shows the Gauss law observable $V_A$, which means the
      observable involving the links 1 and 2, as shown in Figure \ref{fig:GS1}.}
      \label{fig:outputz2}
  \end{figure*}
We first discuss the results for the $\mathbb{Z}_2$ theory on square and
triangular plaquettes, which were simulated on IBM Q Valencia and IBM Q Bogota,
respectively. The results are plotted in Figure \ref{fig:outputz2}. The plots $a$ and $b$
in the top row of the figure show a simulation of a single square plaquette system for
two different couplings: $g=1.0$ and $g=2.0$. We chose IBM Q Valencia for this
simulation because of its T-shaped topology, illustrated in $b$ of Figure
\ref{fig:topology}, which reduced the circuit depth necessary since the ancillary qubit
could be placed at a junction directly connected to three other qubits. There was other
hardware available with better $V_Q$ (32 versus 16 for Valencia), but the topological
advantage of the T-shaped hardware made for better results despite the worse $V_Q$. In
these plots we give the ideal simulator measurement of the Loschmidt probability in addition
to the original (raw) data from the circuit, followed by the readout error correction,
followed by the readout and ZNE error corrections in combination. Here we see that with
both these corrections we are able to get to the correct simulator measurements within
errors.

The next two plots, $c$ and $d$ in the bottom row of Figure \ref{fig:outputz2}
give the results for a $\mathbb{Z}_2$ theory on a triangular plaquette instead. Here a
smaller circuit depth is needed as compared to the square plaquette, so we use IBM Q
Bogota due to its better quantum volume (it has a linear topology, as seen in
$a$ in Figure \ref{fig:topology}). These plots give the time-evolution for the
two states in the $V_A = V_B=V_C = 1$ sector: $\left|000\right\rangle$ and
$\left|111\right\rangle$, and one can see from the simulator lines that their
probabilities always add up to 1. As in the case for the $\mathbb{Z}_2$ theory on the
square plaquette, the error mitigation methods allow for the fully mitigated data to
track the simulator data within error bars. The last plot $e$ in the lower right corner 
of Figure \ref{fig:outputz2} is a measure of how well the circuits for the system on the 
triangular plaquette are producing only states that have $V_A=1$. It shows measurements 
throughout the time evolution of $\left\langle V_A\right\rangle$, and as the simulator 
line shows, ideally it would remain exactly equal to 1 throughout the time evolution. 
The mitigated measurements show how for most time measurements we are able to produce 
$\left\langle V_A\right\rangle =1$ within error bars.

We further note that the circuit depths for the simulations of the $\mathbb{Z}_2$
theory on the square plaquette lead to circuit volumes clearly greater than the
quantum volume $V_Q$ measurements of the quantum hardware ($d=8$, $m=5$ leading to a
circuit volume of $40$ for the square plaquette, whereas $V_Q$ is 16 on IBM Q
Valencia--suggesting a maximum square circuit volume of $16$, with $d=m=4$.) The simple
mitigation techniques employed thus seem to allow us to ``beat'' the quantum volume
limitations for the hardware and get results consistent with the simulator within
errors. For the triangular plaquette on IBM Q Bogota, we have $d=8$, $m=4$, leading to
a circuit volume of $32$, whereas the $V_Q$ of the hardware is $32$, corresponding to a
$d=m=5$ square. It is less clear whether we have exceeded quantum volume limitations for
this simulation, and indeed empirically most Loschmidt probability data seems to meet the IBM
Q threshold of $67\%$ of the ideal amplitude\cite{PhysRevA.100.032328}, but again we
see that our mitigation efforts are successful at restoring the full measurement
values.

 \begin{figure*}
      \centering
      \includegraphics[width=5cm]{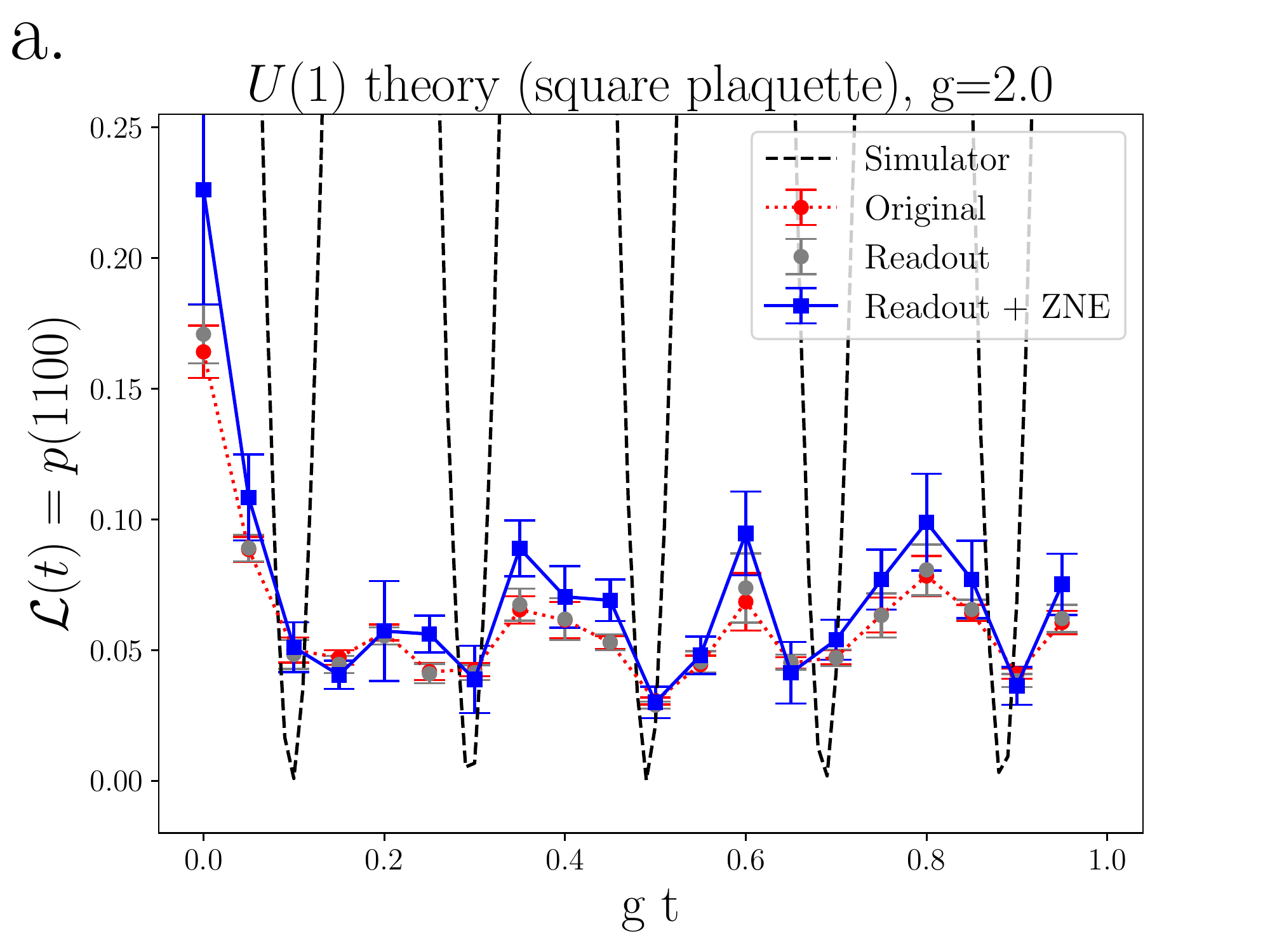}
      \includegraphics[width=5cm]{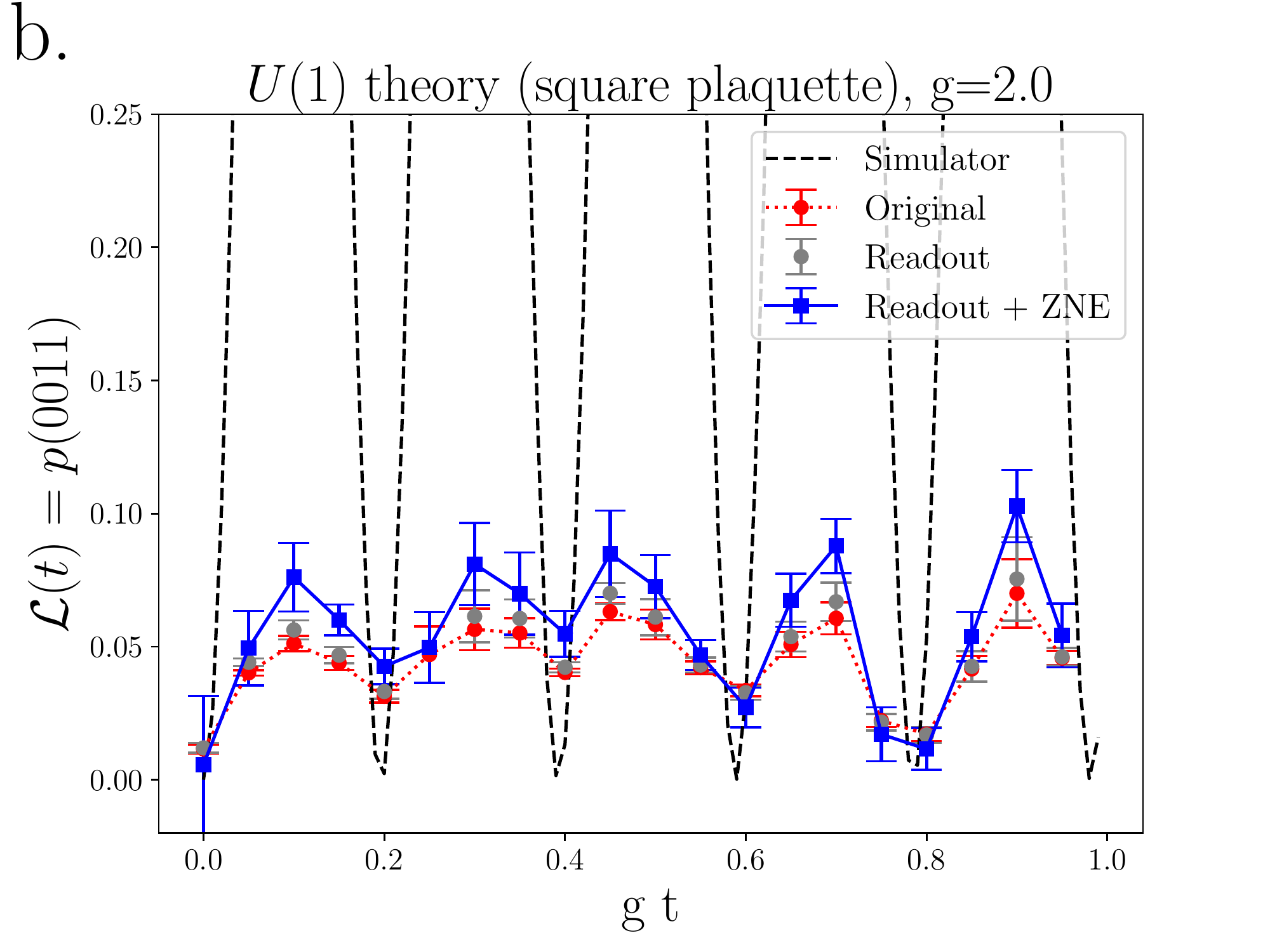}
      \includegraphics[width=5cm]{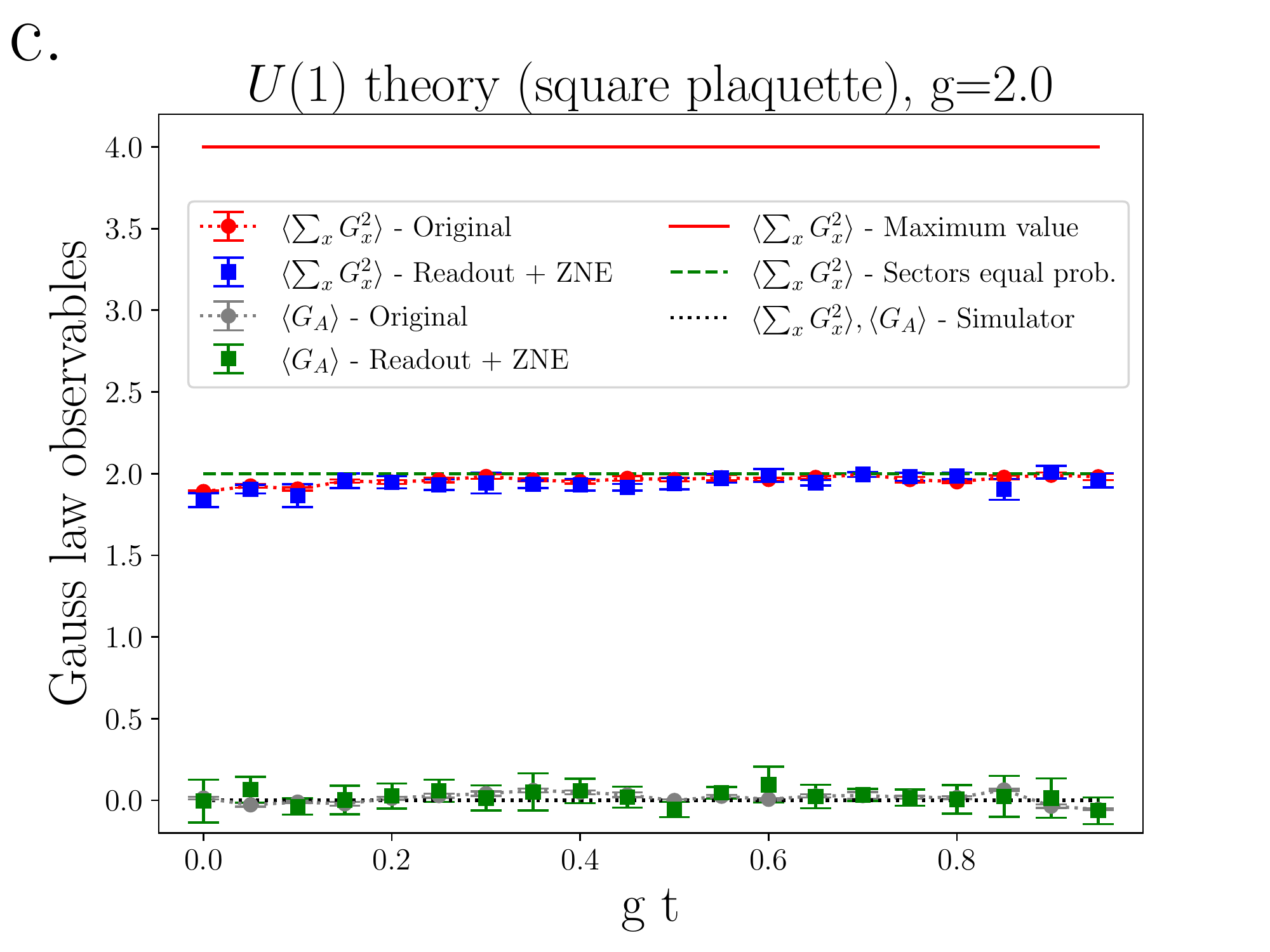}
      \includegraphics[width=5cm]{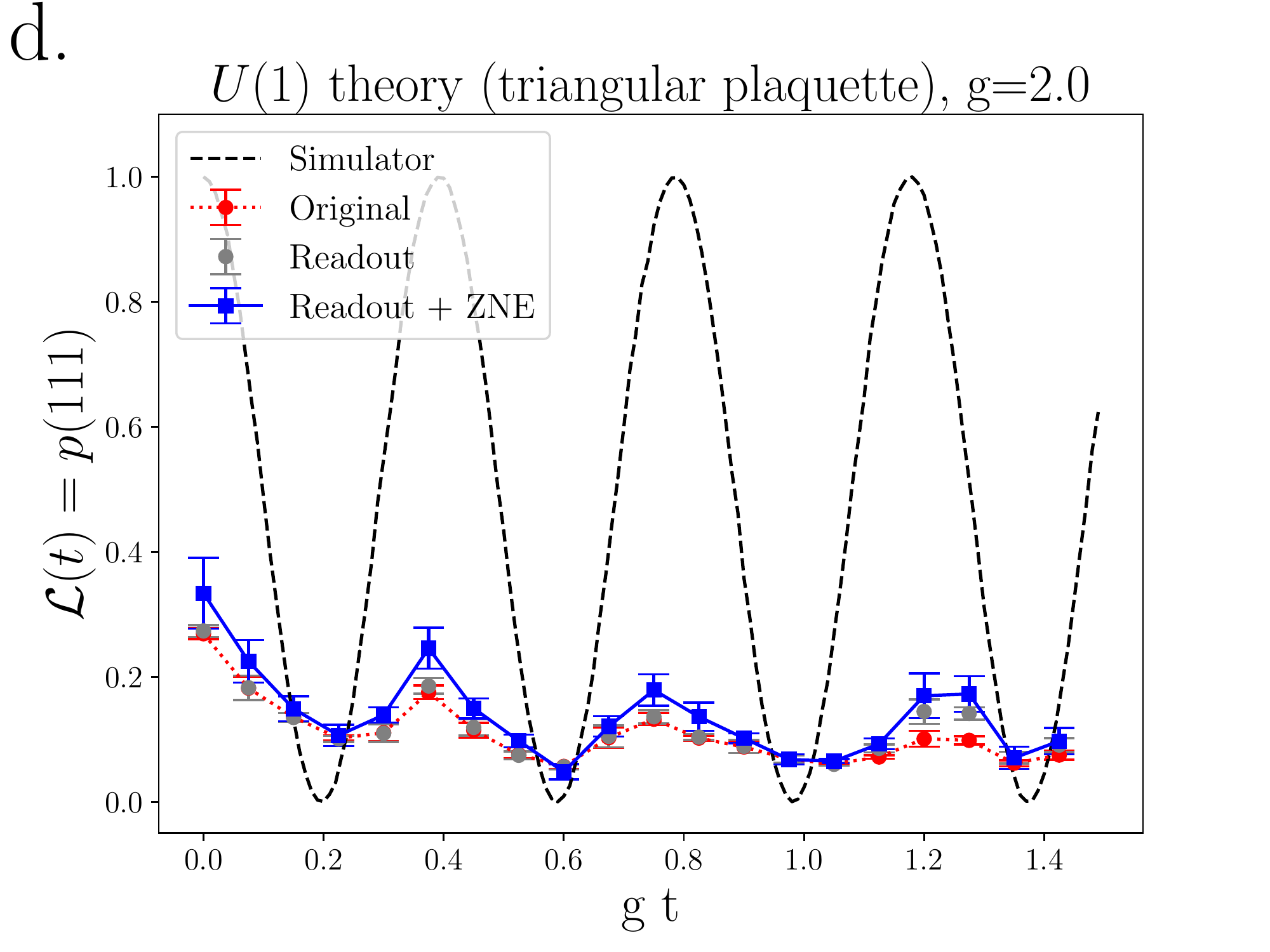}
      \includegraphics[width=5cm]{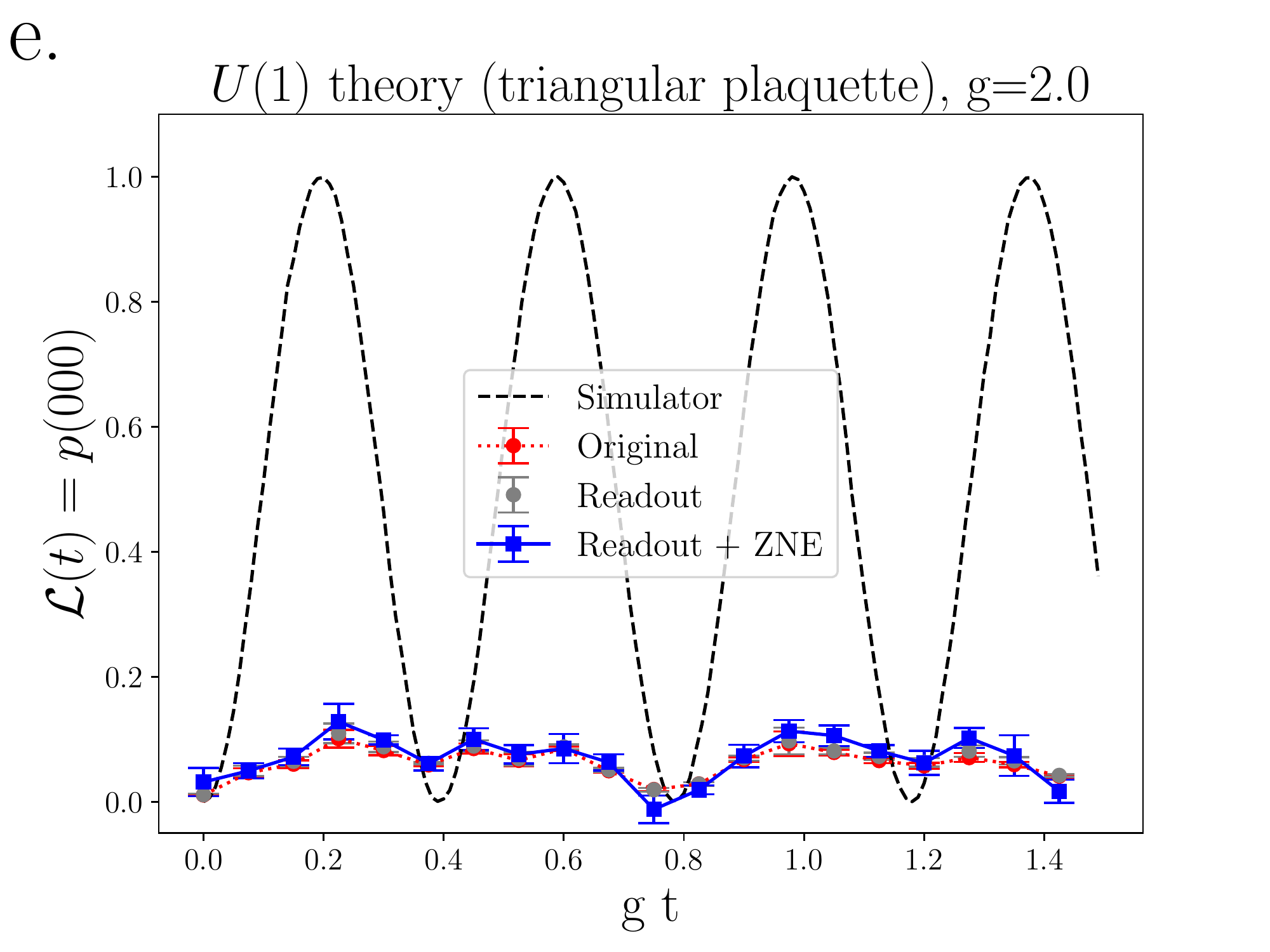}
      \includegraphics[width=5cm]{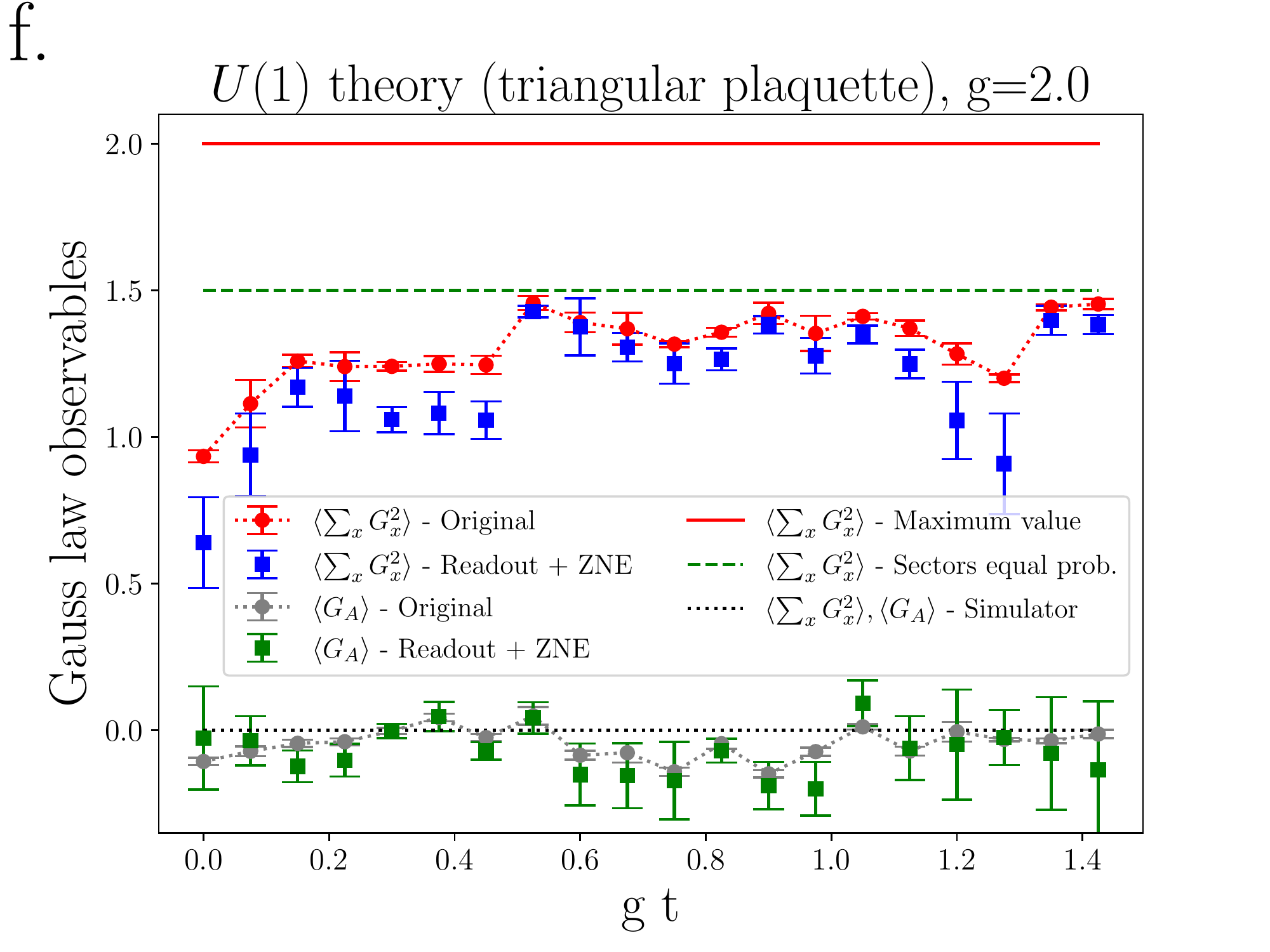}
      \caption{Real-time evolution of the $U(1)$ theory on a single plaquette. The top
      row shows results for the square plaquette on IBM Q Quito hardware, with plots
      $a$ and $b$ showing Loschmidt probability data, and then plot $c$ showing different
      Gauss law observables $G_A$ and $\sum_x G_x^2$, with $G_x$ defined in Equation 
      (\ref{eq:gl}). The bottom row shows results for the triangular plaquette on IBM Q 
      Manila hardware, with again the first two plots $a$ and $b$ showing Loschmidt 
      probability data, and the third showing the $G_A$ and $\sum_x G_x^2$ Gauss law 
      observables.}
     \label{fig:outputu1}
  \end{figure*}

 \subsection{\texorpdfstring{$U(1)$}{U(1)} Theory on Single Plaquettes}
 We next present the data for the $U(1)$ theory on a single square plaquette and a
 single triangular plaquette, which we ran on IBM Q Quito and IBM Q Manila,
 respectively. Similar to the $\mathbb{Z}_2$ case, IBM Q Quito has a T-shaped
 architecture (as seen in $b$ of Figure \ref{fig:topology}) with $V_Q=16$, while
 IBM Q Manila has a linear topology (as seen in $a$ of Figure
 \ref{fig:topology}) with $V_Q=32$. We ran the square plaquette simulation on the
 T-shaped architecture because despite its lower $V_Q$, the topological advantages
 requiring fewer two-qubit gates made for better data. Indeed, we could not get any
 signal at all for the square plaquette $U(1)$ model on current linear-topology IBM Q 
 devices. 
 
 Figure \ref{fig:outputu1} shows the data for the $U(1)$ simulations. The first row
 of plots gives the square plaquette simulation data, with the first two plots $a$ 
 and $b$ showing Loschmidt probability data for the two states $\left|1100\right\rangle$ 
 and $\left|0011\right\rangle$ in the $G_x=0$ sector. Here we are running circuits that 
 have much greater volume than the quantum volume limitations, with $m=5$ and $d=80$, 
 and so we cannot come close to the correct amplitudes of the oscillations (shown by 
 the dashed simulator lines), but we are able to make out some oscillations and see 
 some qualitative similarity between the experimental data and the simulator data. 
 It is clear however that the folding ZNE is unable to improve the accuracy of the 
 data at this level.
 
 The last plot $c$ in the top row is a test of how well the time-evolved system
 stays in the $G_x=0$ sector by measuring two quantities: $G_A$ in particular and 
 then $\sum_x G_x^2$. For both of these quantities we would expect to get zero in 
 the ideal case, and indeed the data for $G_A$ stays quite close to zero. As this 
 is a simple average of $G_A$ however, we cannot rule out that many $G_A$ measurements 
 of $+1$ and $-1$ also exist in roughly equal quantities and are being averaged away, 
 and indeed the leakage seen from the other plots suggests this must be occurring. 
 We can quantify this leakage better by additionally measuring $\sum_x G_x^2$, which 
 ideally should also be equal to $0$ at all times. Here we also plot two lines: one at 
 $4$, which is the maximum value one could possibly get (by staying in the $G_x=\pm 1$ 
 sectors, because the observable would be $4\times (\pm1)^2 = 4$), and one at $2$, which 
 is the value one would get if all sectors were equally represented in the time-evolution 
 (for the sixteen sectors average we would get $(2\times 0 + 12\times 2 + 2\times 4)/16 = 2$). 
 When we look at our experimental data we see that indeed the measurements are quite close 
 to all sectors being equally likely, but they are mostly slightly below that line. 
 This suggests a slight bias toward the $G_x=0$ sector.
 
 The second row of Figure \ref{fig:outputu1} shows the data for the $U(1)$ theory on the 
 triangular plaquette, with the first two plots $d$ and $e$ giving the Loschmidt probability 
 for states $\left|000\right\rangle$ and $\left|111\right\rangle$, which are the two states 
 in the $G_x=0$ sector. Again with $d=4$ and $m=40$ we are likely far past the volume threshold 
 suggested by $V_Q=32$, and indeed the original data never comes close to the maximum amplitudes 
 of $1$ in the oscillations. However, again we are able to make out a qualitative agreement in 
 behavior. We also see a close agreement in the frequency of the oscillations and that ZNE does 
 still incrementally improve the results, unlike in the square plaquette case.
 
 The last plot $f$ in the bottom row again measures $ G_A$ and $\sum_x G_x^2$, and again the $G_A$ 
 observable is mostly close to $0$, but once more, this can be explained by ``leaky'' states in both 
 $G_A=1$ and $G_A=-1$ sectors also being sampled (so long as both the $G_A=1$ and $G_A=-1$ errors 
 are equally likely). We see this more clearly by measuring $\sum_x G_x^2$ as well. Again we show 
 two lines for comparison: the ``maximum value'' line shows the case where we get the largest value 
 for the triangular plaquette, which occurs when $G_x=\pm 1$ for two of the sites and $G_x=0$ for 
 the third site. This results in an average value of $2$, and we see our experimental values are 
 well below that. The second line again shows the value we would get if all sectors in the time 
 evolution were equally likely (obtained by computing $(2\times 0 + 6\times 2) / 8 = 3/2$ for the 
 eight sectors of the triangular plaquette system). Here we see quite clearly that even though 
 our experimental data for $\sum_x G_x^2$ is much larger than $0$, it is also clearly smaller than 
 $3/2$, indicating a clear bias toward the $G_x=0$ sector.

 \begin{figure*}
      \centering
      \includegraphics[width=5cm]{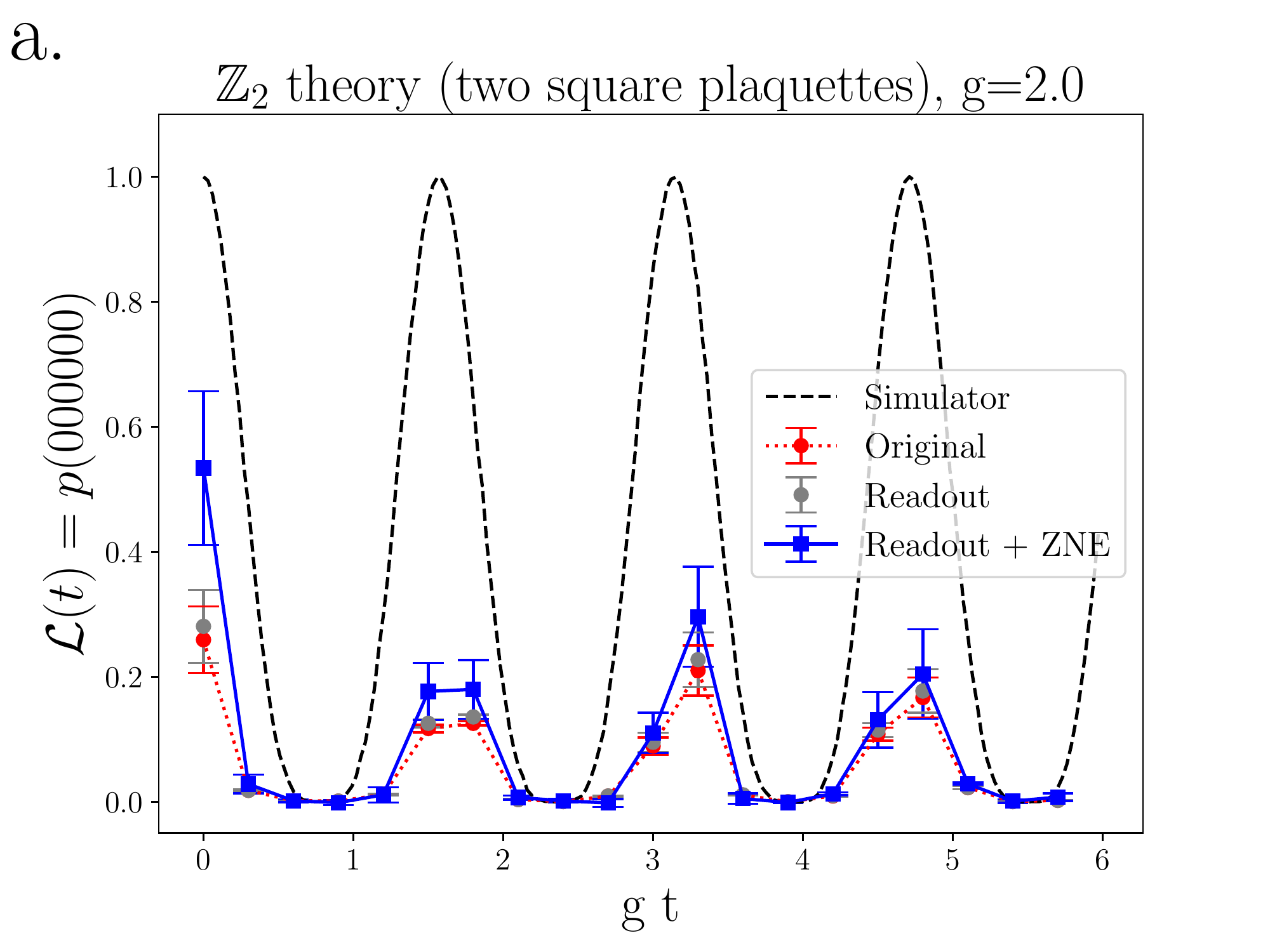}
      \includegraphics[width=5cm]{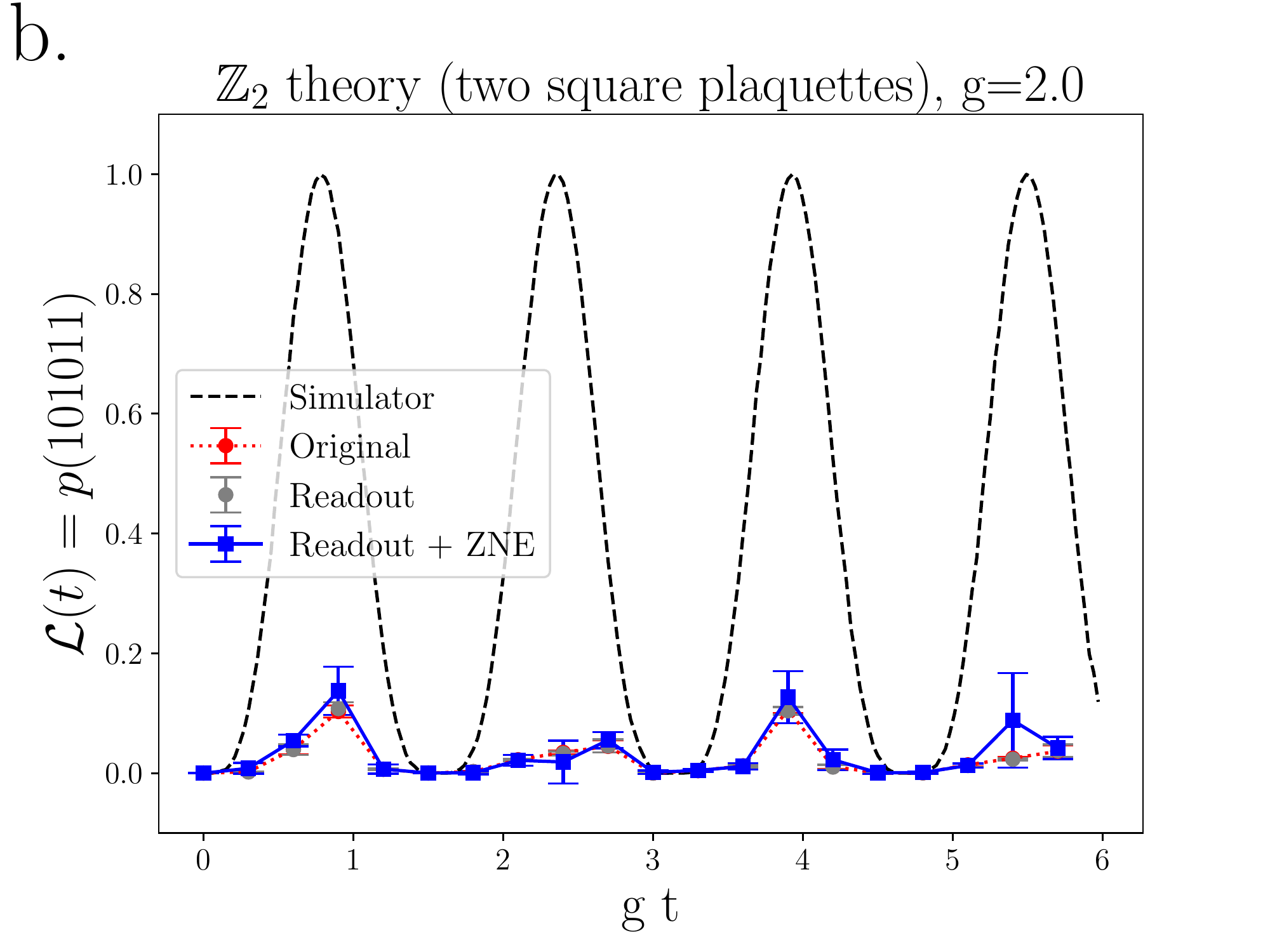}
      \includegraphics[width=5cm]{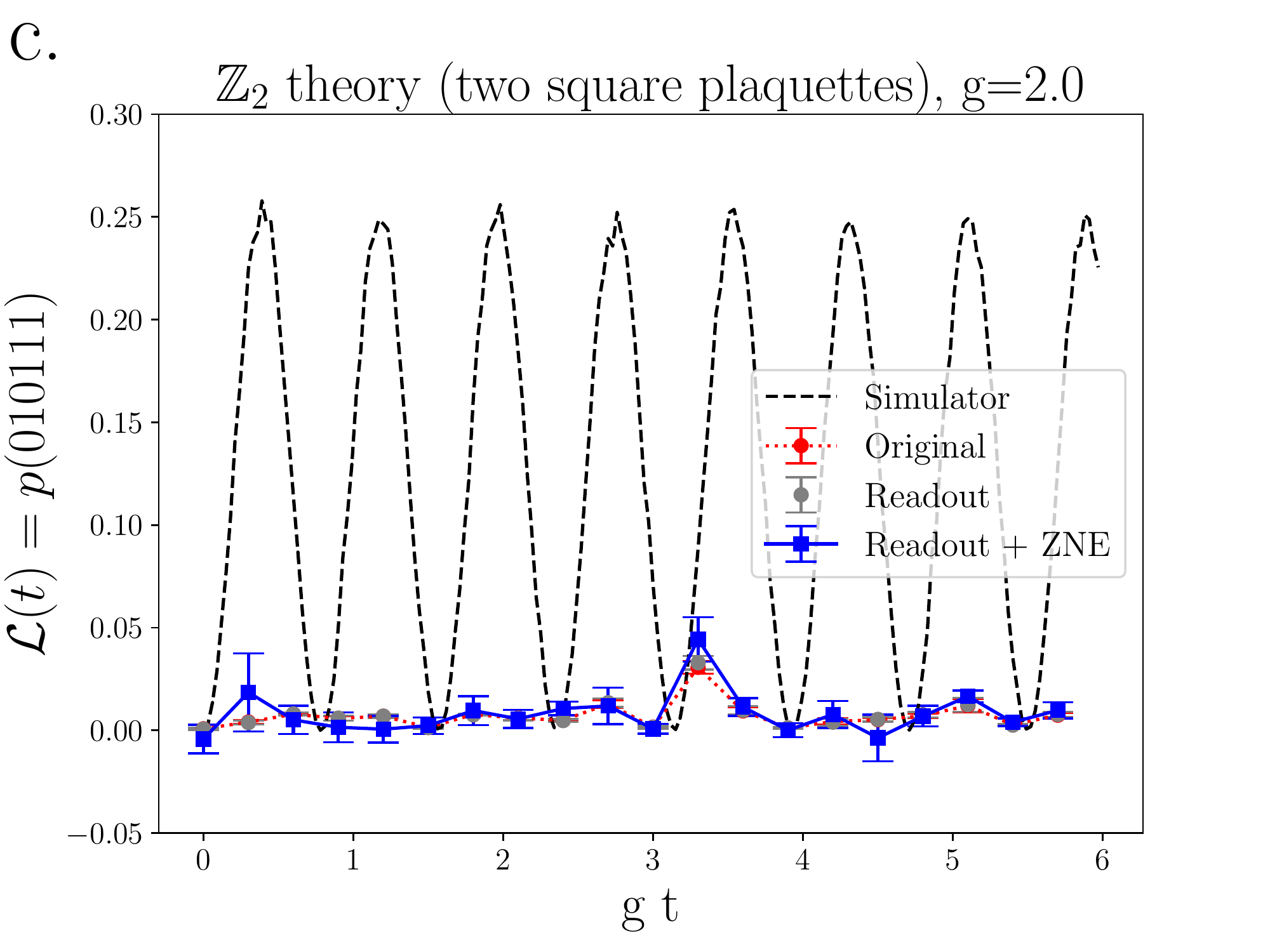}
      \includegraphics[width=5cm]{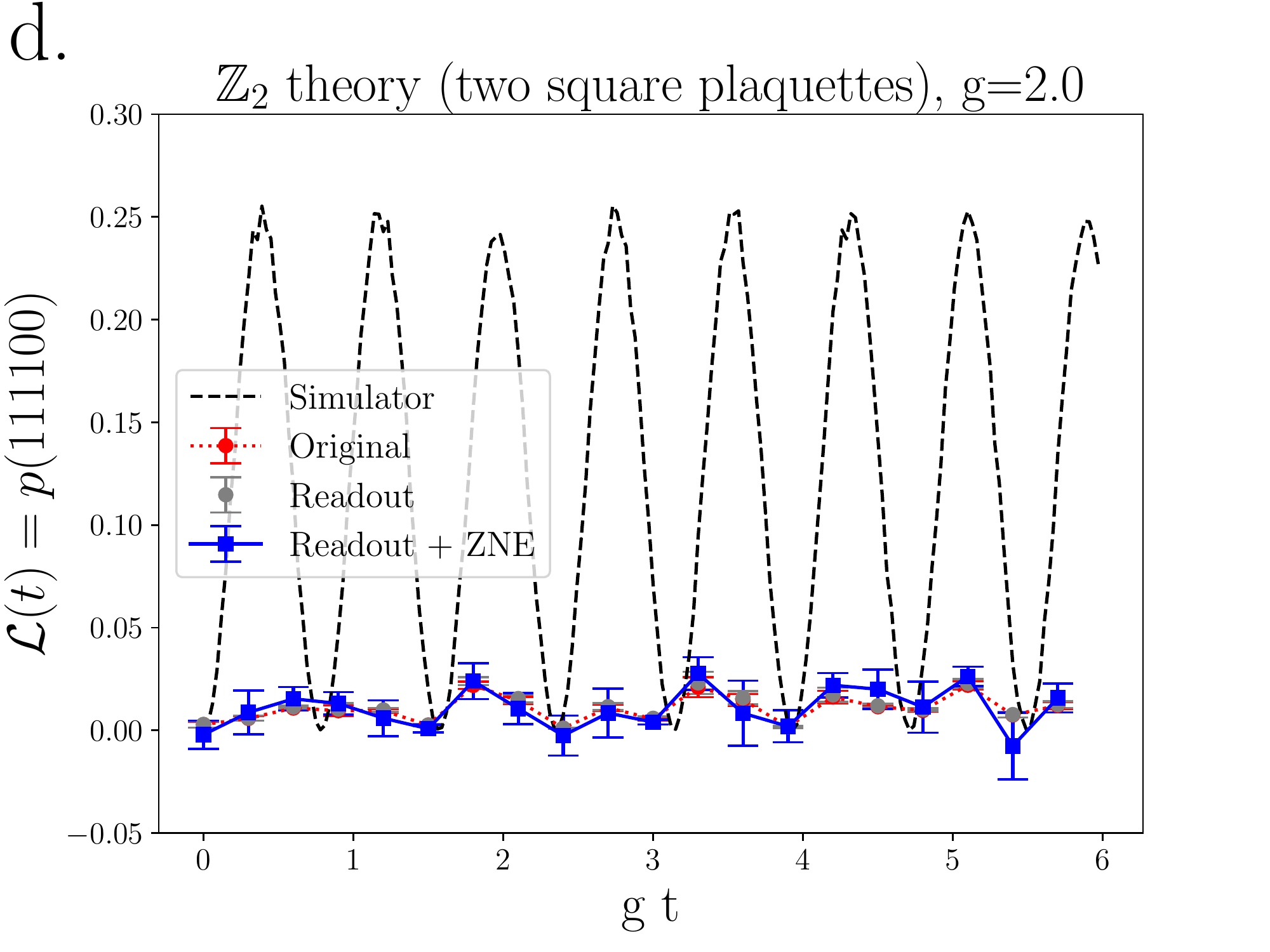}
      \includegraphics[width=5cm]{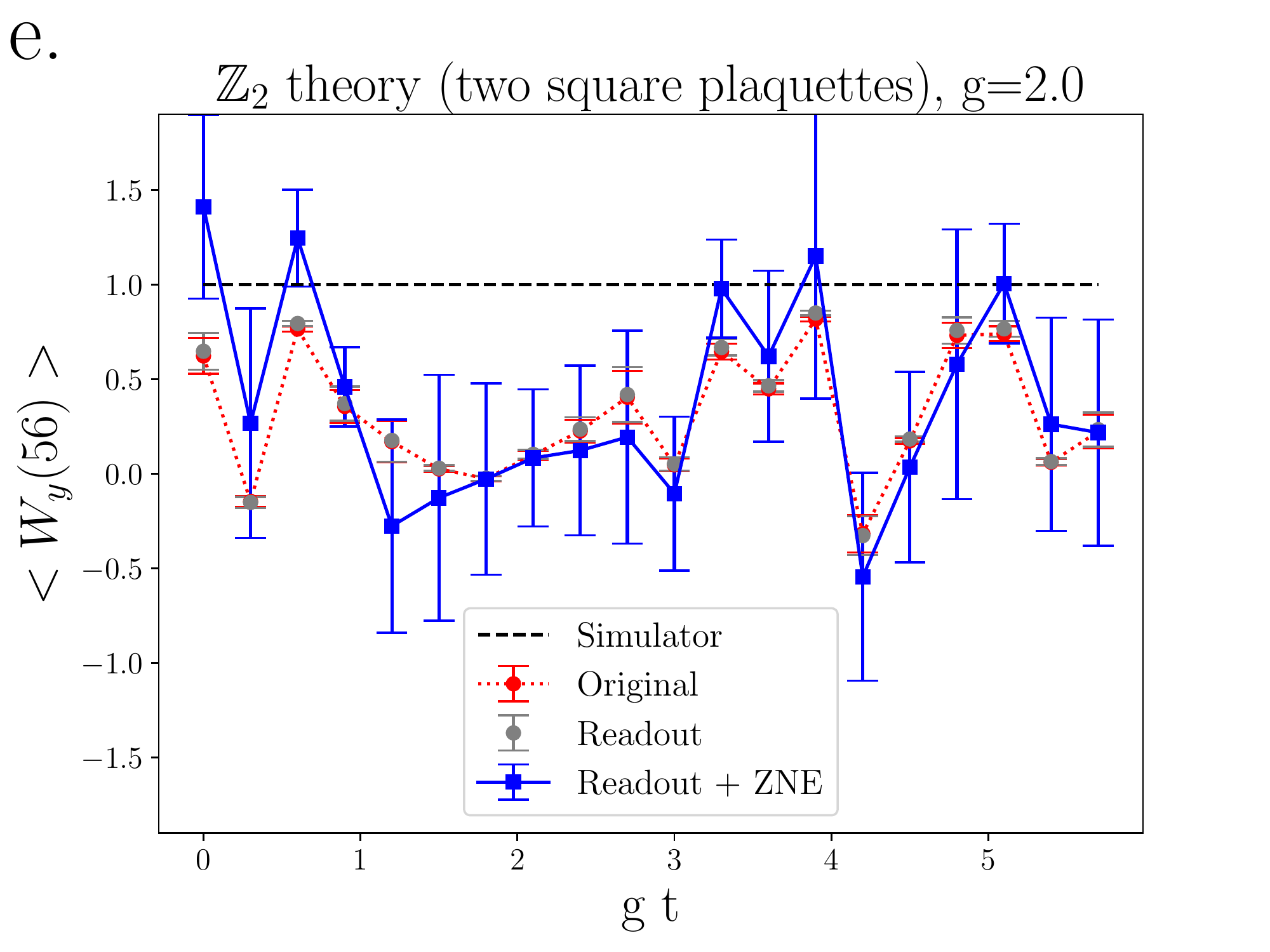}
      \includegraphics[width=5cm]{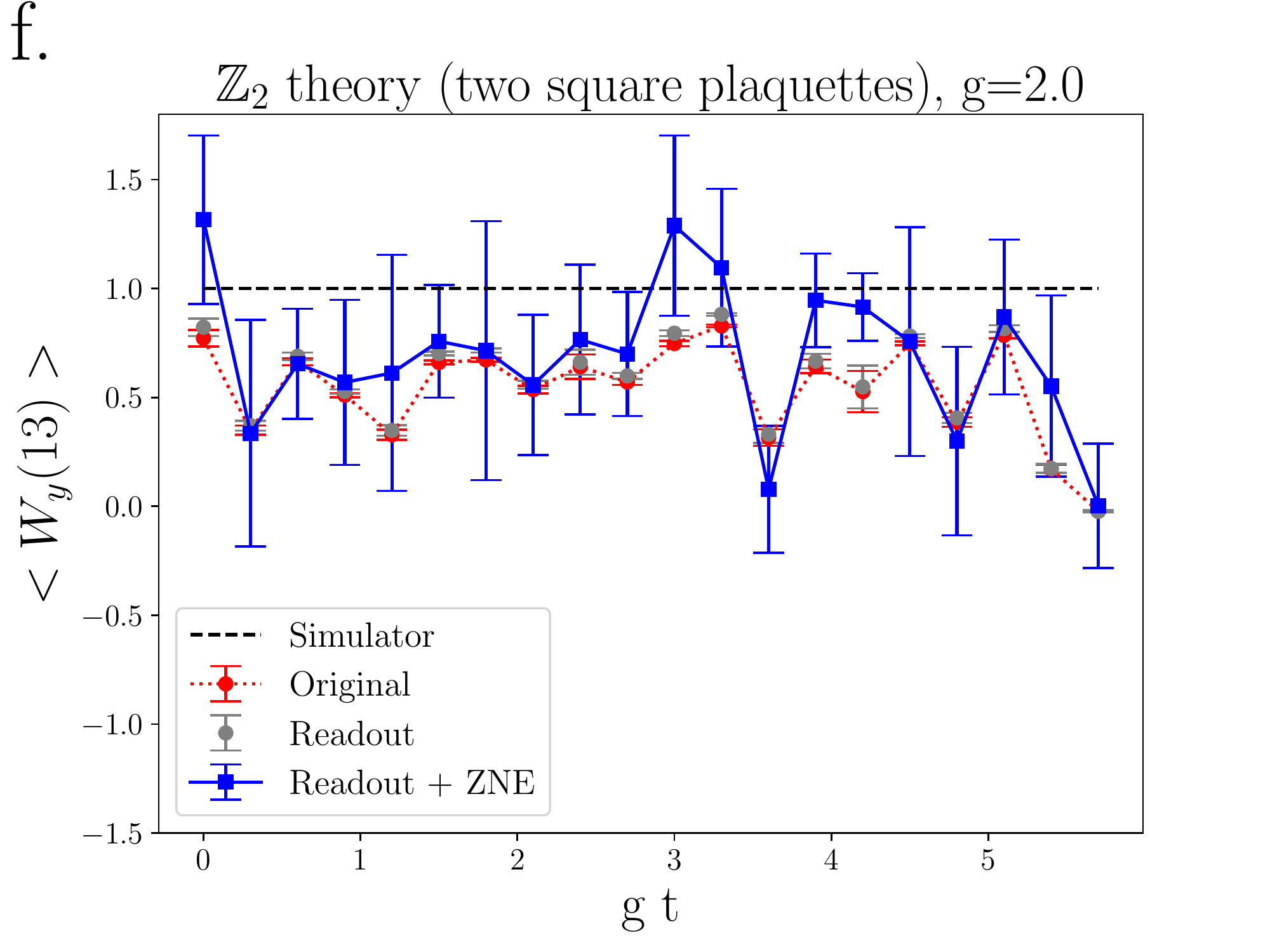}
      \caption{Plots for the two-plaquette $\mathbb{Z}_2$ system, which was run on IBM
      Lagos. The first two plots $a$ and $b$ give the Loschmidt probabilities for states
      $\left|000000\right\rangle$ and $\left|101011\right\rangle$ which oscillate between 
      $0$ and $1$, and the next two plots $c$ and $d$ give the Loschmidt probabilities 
      for states $\left|010111\right\rangle$ and $\left|111100\right\rangle$, which 
      oscillate between $0$ and $0.25$. The last two plots $e$ and $f$ are for the 
      winding number observables in the $y$-direction, the first involving links 5 and 6, 
      and the second involving links 1 and 3, as defined in Figure \ref{fig:2plaq}.}
      \label{fig:output2p}
  \end{figure*}

 \subsection{\texorpdfstring{$\mathbb{Z}_2$}{Z(2)} Theory: Two-plaquette System}
 Finally we turn to the time-evolution of the $\mathbb{Z}_2$ theory on the
 two-square-plaquette system, whose ideal behavior was shown in Figure
 \ref{fig:2plaqSimulator}, where we see that if the system's initial state is in the 
 sector where $G_x=1$,
 the system's evolution involves only the four states that fall into that sector. As
 illustrated by Figure \ref{fig:2plaq}, we are using periodic boundary conditions, and
 so there are six distinct links in the two-square-plaquette system. With the addition
 of an ancillary qubit, that brings us to seven qubits minimum for our simulation, and
 so we used the seven-qubit IBM Lagos device to obtain real-time dynamics data. 
 
 Figure \ref{fig:output2p} gives the results for the simulation, with the first four
 plots $a$-$d$ giving Loschmidt probability data for the four states in the $G_x=1$ sector; which we
 label $\left|000000\right\rangle$, $\left|101011\right\rangle$,
 $\left|010111\right\rangle$, and $\left|111100\right\rangle$ in reference to the
 numbered links in Figure \ref{fig:2plaq}. The $V_Q=32$ for IBM Lagos tells us that
 the maximum square circuit meeting the accuracy threshold is $5\times 5$. Comparing that to
 the two plaquette system circuit requirement with $m=7, d=48$ indicates that we are 
 way beyond the quantum volume limit.
 However, especially for the states $\left|000000\right\rangle$ and
 $\left|101011\right\rangle$; where the simulator shows us the maximum amplitude goes
 up to $1$, we are able to see qualitative agreement and the readout error and ZNE
 error corrections do provide incremental improvements to the results.
 
 The last two plots $e$ and $f$ give data for the winding number observable $W_y$, defined in
 Equation \ref{eq:wind}. As noted from before, the winding number in the $y$-direction
 can be measured using links 1 and 3 as well as links 5 and 6, and in each case the result
 should be the same throughout the time-evolution for the initial conditions that we
 chose: $W_y=1$. Indeed when we take the data and use ZNE, we do see a bias in the data
 closer to $+1$ than $-1$ for both $W_y$ observables. As discussed in the section \ref{sec:models},
 the winding number is a topological quantity, which is dependent on how the spins along
 a line spanning the entire system behaves. It is thus expected that this quantity could
 be robust against decoherence noise. In fact, our results here qualitatively confirm this,
 since we see that the winding number expectation value stays close to the winding number
 sector that the initial state belonged to. Of course, one needs to verify this on larger
 circuits.

\section{Conclusions}
\label{sec:conc}
In this paper we have explored the possibilities for real-time simulations of
plaquette theories on current NISQ hardware, including theories with $\mathbb{Z}_2$
symmetries as well as the $U(1)$ symmetry, which is of particular interest from the QED
perspective. We find that for the $\mathbb{Z}_2$ single plaquette models, we can 
successfully overcome quantum volume, $V_Q$, limitations with the error mitigation schemes of readout error 
mitigation as well as ZNE through circuit folding. In cases where the
circuit significantly exceeds the quantum volume, such as the cases of the two-plaquette
$\mathbb{Z}_Z$ model and the $U(1)$ models,
the error mitigation does not have a significant effect on the results. However, since the error mitigation 
techniques used are hardware agnostic, this has promising implications for NISQ devices 
in general, rather than only on IBM Q devices. Even in cases where we cannot overcome 
$V_Q$ limitations, we are still able to see qualitative signals of the real-time dynamics 
for circuits that are many times deeper than the $V_Q$  measurements for the hardware. 
We have seen that topology is also an important consideration for quantum simulations 
with superconducting qubits in particular, and found significant quantitative advantages 
in choosing the best topology for each experiment. 

Future improvements specific to superconducting qubits would involve using 
pulse control for ZNE rather than folding, as well as denser data points to capture 
the time evolution for a plaquette model. Additionally, future work could involve 
simulating the real-time dynamics of non-Abelian plaquette models. Another immediate 
attempt would be to use different encoding strategies already with the microscopic model. 
For example, the $U(1)$ or the $\mathbb{Z}_2$ models can be represented in terms of dual 
height variables in 2-spatial dimensions, which already removes much of the gauge 
non-invariant states. Formulating quantum circuits on the dualized versions of such 
models would enable bigger lattices to be realized on quantum circuits \cite{banerjee2021nematic}. 
Similarly, the use of rishons allows a gauge invariant formulation of several 
non-Abelian gauge theories such as the aforementioned $SO(3)$-symmetric model, 
which can then be used to construct quantum circuits on NISQ devices \cite{Brower_1999, Rico_2018}.

\section*{Acknowledgments}
We would like to thank Sebastian Hassinger, IBM, and Roger Melko, for arranging an
Academic Research Program agreement for us with the IBM Q Experience. We would also
like to thank the Unitary Fund for additional research account access. Thanks are
due to Lukas Rammelm\"uller for providing valuable suggestions on an earlier draft.
Research of EH at the Perimeter Institute is supported in part by the Government of 
Canada through the Department of Innovation, Science and Economic Development and 
by the Province of Ontario through the Ministry of Colleges and Universities.

The source code for our experiments is available at 
\href{https://github.com/mgarciav88/plaquette-models}{https://github.com/mgarciav88/plaquette-models}.

\bibliography{ref}
%\newpage
\appendix

\section{Proof of circuit identity}\label{ap: circuitProof}
Let us label
$S^3_{\rm N} = \sum_{j=1}^N \sigma^3_j$. Using this definition, we want to
prove
\begin{equation}
\begin{aligned}
  U_{\rm S,A}  (t)
  = &\exp \left[ i \frac{\phi}{2} \sigma^3_{\rm A} S^3_{\rm N} \right]  
  \exp \left[ i g t \sigma^1_{\rm A}\right] \\ &\qquad \times 
  \exp \left[- i \frac{\phi}{2} \sigma^3_{\rm A} S^3_{\rm N} \right].
   \label{eq:qgate2}
   \end{aligned}
\end{equation}
The physics behind the implementation is that the real-time evolution is performed 
on a single qubit, called the ancillary qubit. However, before and after, the ancillary 
qubit is entangled with the ${\rm N}$ qubits, so that the required dynamics is also 
induced on them.

To prove the relation, we first note,
\begin{equation}
    \sigma^1_{\rm A} \exp \left[ i \frac{\phi}{2} \sigma^3_{\rm A} S^3_{\rm N} \right] = 
    \exp \left[- i \frac{\phi}{2} \sigma^3_{\rm A} S^3_{\rm N} \right]
    \sigma^1_{\rm A}.
    \label{eq:sigma_exp}
\end{equation}
This relation can be obtained by expanding the exponential and noting that
$\sigma_{\rm A}$ commutes with all the other $\sigma_{j}$, and satisfies the
following anti-commutation relations
\begin{equation}
    \{ \sigma^\alpha_{\rm A},  \sigma^\beta_{\rm A} \} = 2 \delta^{\alpha \beta}.
\end{equation}
We then commute the $\sigma^1_{\rm A}$ across at the expense of a negative sign
in the series, and then re-exponentiating it, proves the relation \ref{eq:sigma_exp}.

Thus:
\begin{equation}
\begin{split}
        U_{\rm S,A} (t) & = \exp \left[ i \frac{\phi}{2} \sigma^3_{\rm A} S^3_{\rm N} \right]  
        \left[ {\rm cos}(g t) -i {\rm sin} (g t) \sigma^1_{\rm A}\right] \\
        &\qquad\times\exp \left[- i \frac{\phi}{2} \sigma^3_{\rm A} S^3_{\rm N} \right] \\
        & = {\rm cos}(g t) - i {\rm sin} (g t) \sigma^1_{\rm A} \exp 
        \left[- i \phi \sigma^3_{\rm A} S^3_{\rm N} \right]
\end{split}
\label{eq:qgate3}
\end{equation}
Next, we note that
 \begin{equation}
\begin{aligned}
 (\sigma^1_{\rm A} & \exp \left[- i \phi \sigma^3_{\rm A} S^3_{\rm N} \right])^n\\
 & \qquad\qquad = \left\{ \begin{array}{cc} 1,&\; n\;{\rm even} \\
  \sigma^1_{\rm A} \exp \left[- i \phi \sigma^3_{\rm A} S^3_{\rm N} \right],&\;
    n\;{\rm odd} \end{array} \right.
\end{aligned}
\label{eq:inter2}
\end{equation}
For $n=2$,
\begin{equation*}
\begin{aligned}
  & (\sigma^1_{\rm A} \exp \left[- i \phi \sigma^3_{\rm A} S^3_{\rm N} \right])
   (\sigma^1_{\rm A} \exp \left[- i \phi \sigma^3_{\rm A} S^3_{\rm N} \right]) \\
   &\qquad\qquad = (\sigma^1_{\rm A})^2 \exp \left[ i \phi \sigma^3_{\rm A} S^3_{\rm N} \right]
   \exp \left[-i \phi \sigma^3_{\rm A} S^3_{\rm N} \right] = 1.
   \end{aligned}
\end{equation*}
Similarly, for $n=3$,
\begin{align*}
   &
   (\sigma^1_{\rm A} \exp \left[- i \phi \sigma^3_{\rm A} S^3_{\rm N} \right])
   (\sigma^1_{\rm A} \exp \left[- i \phi \sigma^3_{\rm A} S^3_{\rm N} \right])
   \\& \qquad\qquad\times (\sigma^1_{\rm A} \exp \left[- i \phi \sigma^3_{\rm A} S^3_{\rm N} \right]) \\
   & =
   (\sigma^1_{\rm A})^2 \exp \left[i \phi \sigma^3_{\rm A} S^3_{\rm N} \right]
   \exp \left[-i \phi \sigma^3_{\rm A} S^3_{\rm N} \right] \\
   &\qquad \qquad\times (\sigma^1_{\rm A} \exp \left[- i \phi \sigma^3_{\rm A} S^3_{\rm N} \right]) \\
   & =
   (\sigma^1_{\rm A} \exp \left[- i \phi \sigma^3_{\rm A} S^3_{\rm N} \right]).
\end{align*}
and hence the case for general $n$ follows. Then, we can re-exponentiate to give
\begin{equation}
\begin{split}
    U_{\rm S,A} &(t)  = {\rm cos}(g t) -i {\rm sin} (g t) \sigma^1_{\rm A}
    \exp \left[- i \phi \sigma^3_{\rm A} S^3_{\rm N} \right] \\
    & = \exp \left[- i g t \sigma^1_{\rm A} \exp\{- i \phi \sigma^3_{\rm A} S^3_{\rm N} \} \right] \\
     & = \exp \left[- i g t \sigma^1_{\rm A} \prod_{j=1}^N
     ( {\rm cos}(\phi) - i \sigma^3_{\rm A} \sigma^3_{j} {\rm sin}(\phi) ) \right]
\end{split}
\end{equation}
For $\phi = \frac{\pi}{2}$ (and hence the factor $\frac{\pi}{4}$ in Equation \ref{eq:qgate1}, we get
\begin{equation}
   U_{\rm S,A} (t) = \exp \left[- i g t \sigma^1_{\rm A}
   \prod_{j=1}^N (- i \sigma^3_{\rm A} \sigma^3_{j})\right]
   \label{eq:qgate4}
\end{equation}
For $N = 1$,
\begin{equation}
\begin{aligned}
   U_{\rm S,A} (t) &=  \exp \left[- i g t \sigma^1_{\rm A} (-i \sigma^3_{\rm A} \sigma^3_{1} ) \right] \\ 
   &= \exp \left[i g t \sigma^2_{\rm A} \sigma^3_1 \right]
    \label{eq:time_evolN1}
    \end{aligned}
\end{equation}
For $N = 2$,
\begin{equation}
\begin{aligned}
   U_{\rm S,A} (t) &=  \exp \left[- i g t \sigma^1_{\rm A} (-i \sigma^3_{\rm A} \sigma^3_{1}) 
   (-i \sigma^3_{\rm A} \sigma^3_{2})\right] \\ 
   &= \exp \left[i g t \sigma^1_{\rm A} \sigma^3_1 \sigma^3_2\right]
    \label{eq:time_evolN2}
    \end{aligned}
\end{equation}
For $N = 3$,
\begin{equation}
\begin{aligned}
   U_{\rm S,A} (t) &=  \exp \left[(-i)^4 g t \sigma^1_{\rm A}
   (\sigma^3_{\rm A} \sigma^3_{1}) (\sigma^3_{\rm A} \sigma^3_{2})
   (\sigma^3_{\rm A} \sigma^3_{3}) \right] \\ &=
   \exp \left[-i g t \sigma^2_{\rm A} \sigma^3_1 \sigma^3_2 \sigma^3_3\right]
   \label{eq:time_evolN3}
   \end{aligned}
\end{equation}
For $N = 4$,
\begin{equation}
\begin{aligned}
   &U_{\rm S,A} (t)\\ &\;=  \exp \left[(-i)^5 g t \sigma^1_{\rm A}
   (\sigma^3_{\rm A} \sigma^3_{1}) (\sigma^3_{\rm A} \sigma^3_{2})
   (\sigma^3_{\rm A} \sigma^3_{3}) (\sigma^3_{\rm A} \sigma^3_{4}) \right] \\ &\;=
   \exp \left[-i g t \sigma^1_{\rm A} \sigma^3_1 \sigma^3_2 \sigma^3_3 \sigma^3_4 \right]
    \label{eq:time_evolN4}
    \end{aligned}
\end{equation}
From $N=5$, the pattern repeats itself.
\subsection{Two-qubit gate combination identity}
The previous identity shows us that we need to find a way to express
\begin{equation}
   e^{-i(\phi/2) \sigma^z_1 \sigma^z_i}
\end{equation}
using two-qubit gates. Using the basis $00, 01, 10, 11$, this two-qubit gate is given by
\begin{equation}
  \left( \begin{array}{cccc} e^{-i\phi/2} & 0 & 0 & 0\\
   0 & e^{i\phi / 2} & 0 & 0 \\
   0 & 0 & e^{i\phi / 2} & 0 \\
   0 & 0 & 0 & e^{-i \phi/2}\end{array}\right) .
   \label{eq:basis}
\end{equation}
We can get this gate, up to a constant, using
\begin{equation}
\begin{aligned}
   e^{-i\phi/2 \sigma^z_1}& =   \left( \begin{array}{cccc} e^{-i\phi/2} & 0 & 0 & 0\\
   0 & e^{-i\phi / 2} & 0 & 0 \\
   0 & 0 & e^{i\phi / 2} & 0 \\
   0 & 0 & 0 & e^{i \phi/2}\end{array}\right), \\   e^{-i\phi/2 \sigma^z_i} &=   
   \left( \begin{array}{cccc} e^{i\phi/2} & 0 & 0 & 0\\
   0 & e^{-i\phi / 2} & 0 & 0 \\
   0 & 0 & e^{i\phi / 2} & 0 \\
   0 & 0 & 0 & e^{-i \phi/2}\end{array}\right),\\
   CP_{1i}(\phi)& =  \left( \begin{array}{cccc} 1& 0 & 0 & 0\\
   0 &1 & 0 & 0 \\
   0 & 0 & 1 & 0 \\
   0 & 0 & 0 & e^{-i2 \phi}\end{array}\right)
   \end{aligned}
\end{equation}
The product then is
\begin{equation}
\begin{aligned}
   e^{-i\phi/2 \sigma^z_1} e^{-i\phi/2 \sigma^z_i}& CP_{1i}\\ & =   
   \left( \begin{array}{cccc} e^{-i\phi} & 0 & 0 & 0\\
   0 & 1 & 0 & 0 \\
   0 & 0 & 1 & 0 \\
   0 & 0 & 0 & e^{-i \phi}\end{array}\right),
   \end{aligned}
\end{equation}
which, up to a factor of $e^{i\phi/2}$, is the same as (\ref{eq:basis}).

\section{Solution of the 2-plaquette system}\label{ap:twoPlaquette}
\begin{figure}
      \centering
      \includegraphics[scale=0.4]{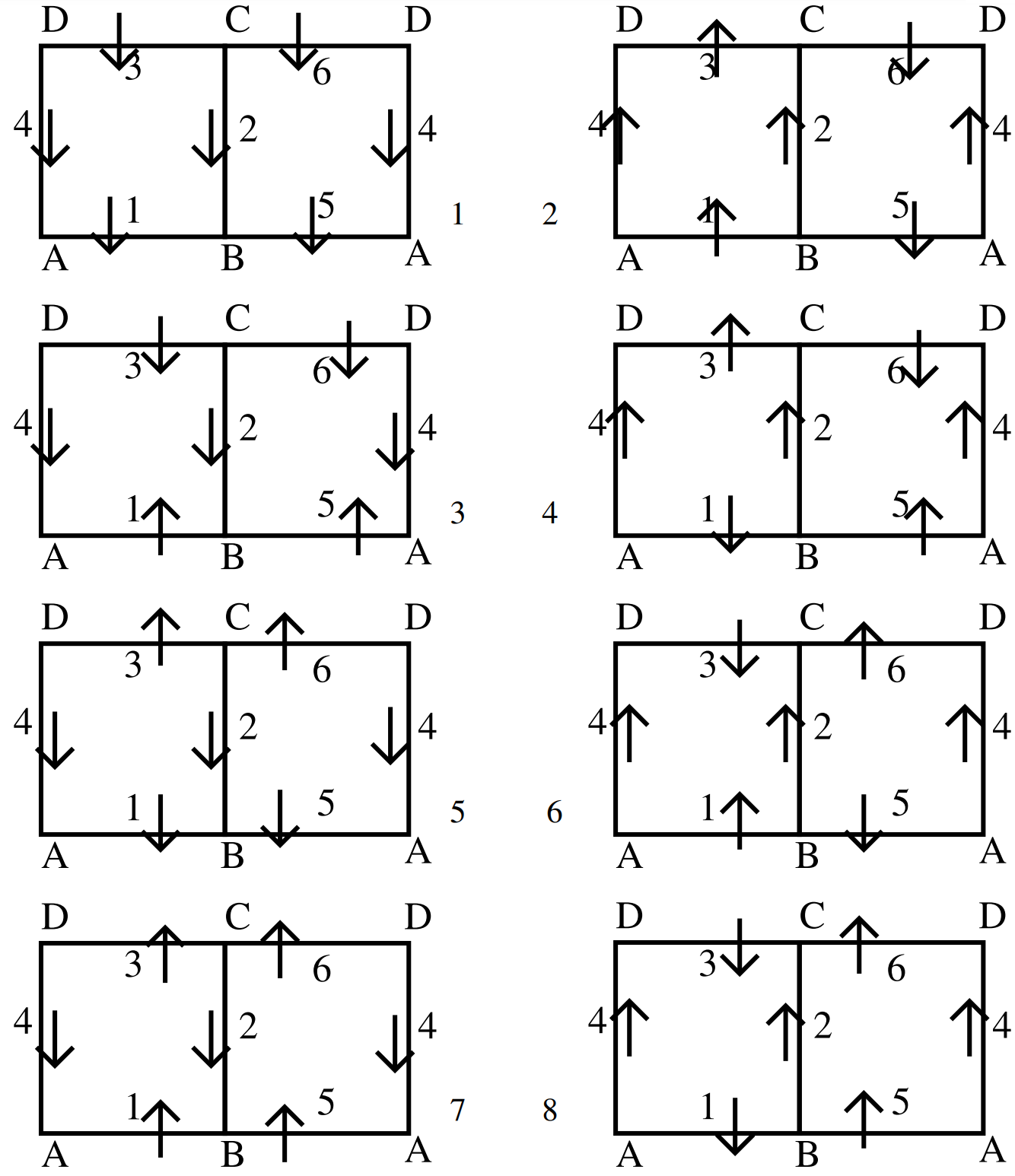}
      \caption{Basis states for the 2-plaquette $Z(2)$ lattice gauge theory. The up- 
      and the down-spins are denoted in the x-basis.}
      \label{fig:cartoonZ22plaq}
  \end{figure}

  There are 8 basis states which obey the Gauss law (see Equation
  ~\ref{eq:Z2_2plaquette}), and we label them as follows (working in the $\Sx{}$
  computational basis)
 \begin{equation}
 \begin{split}
\ket{e_1}&=\ket{000000};~~~\ket{e_2}=\ket{001111};~~~\ket{e_3}=\ket{010001}; \\
\ket{e_4}&=\ket{011110};~~~\ket{e_5}=\ket{100100};~~~\ket{e_6}=\ket{101011};\\ 
\ket{e_7}&=\ket{110101};~~~\ket{e_8}=\ket{111010}; 
\end{split}
 \label{eq:states_2plaq}
 \end{equation}
  In this notation the states are represented by a string of 0s and 1s, the former
  denoting a spin-down and the latter a spin-up in the $\Sx{}$-basis. The six bits
  refer to the spins on the bonds labelled as 6,5,4,3,2,1 in the Fig
  \ref{fig:2plaq}. As a concrete example the basis state $\ket{e_8}$ represents the
  spins on the links 2, 4, 5, and 6 with $\Sx{} = +1$; and those at 1 and 3
  with $\Sx{} = -1$. The basis states are pictorially denoted in Figure \ref{fig:cartoonZ22plaq}. 
  On closer inspection it is clear that the basis states have another quantum
  number: these are the winding numbers of the $Z(2)$ strings along the lattice 
  x- and y-directions respectively. The operators corresponding to these are simply
  the product of the $\Sx{}$ operators along a line which cuts the plaquettes
  horizontally and vertically respectively. For our case, the expressions
  for the operators are 
  \begin{align}
      W_x & = \sigma^x_4 \sigma^x_2;~~~W_y  = \sigma^x_1 \sigma^x_3;~~~W_y  = \sigma^x_5 \sigma^x_6. 
  \end{align}
  The last two expressions for $W_y$ are actually the same as can be seen by using
  the Gauss law for the sites. The basis states $\ket{e_1}, \ket{e_2}, \ket{e_7},
  \ket{e_8}$ are in the sector $(1,1)$ while the rest $\ket{e_3}, \ket{e_4},
  \ket{e_5}, \ket{e_6}$ are in the sector $(1,-1)$. These sectors do not mix under
  a unitary (Hamiltonian) evolution.
 
   The Hamiltonian has two terms: the first one will flip the spins 1,2,3, and 4
   (since it is a $\Sz{}$), and the second one flips the spins 5,4,6, and 2. After
   this, one can compare the flipped state to the original one to find out the
   matrix elements. This gives the following Hamiltonian matrix: \\
\begin{align}
H = \begin{bmatrix}
 0 & -1 & 0 & 0 & 0 & 0 & 0 & -1 \\
-1 & 0 & 0 & 0 & 0 & 0 & -1 & 0 \\
 0 & 0 & 0 & -1 & 0 & -1 & 0 & 0 \\
 0 & 0 & -1 & 0 & -1 & 0 & 0 & 0 \\
 0 & 0 & 0 & -1 & 0 & -1 & 0 & 0 \\
 0 & 0 & -1 & 0 & -1 & 0 & 0 & 0 \\
 0 & -1 & 0 & 0 & 0 & 0 & 0 & -1 \\
 -1 & 0 & 0 & 0 & 0 & 0 & -1 & 0 \\
\end{bmatrix}
\label{eq:2plaqZ2H}
\end{align}
  We implement a real-time quench: we start from a basis state in each (winding)
  sector and then compute the probability of finding the evolved wavefunction in
  the initial starting state. With a knowledge of the eigenvectors and eigenvalues
  of the matrix in Equation~\ref{eq:2plaqZ2H}, which we denote as $\ket{\psi_{\rm n}}$,
  and with $\ket{\psi_{\rm i}}$  as the initial state
 \begin{equation}
 \begin{split}
 \ket{\psi(t)} &= \exp{-i H t} \ket{\psi_{\rm i}}, \\
  &= \exp{-i H t} \sum_{\rm n} \ket{\psi_{\rm n}} 
  \braket{\psi_{\rm n} | \psi_{\rm i}} \\
  &= \sum_{\rm n} \exp{-i E_{\rm n} t} c_{\rm n} \ket{\psi_{\rm n}} \\
 \braket{c_{\rm i} | \psi(t) } &= \sum_{\rm n} |c_{\rm n}|^2 \exp{-i E_{\rm n} t}
 \end{split}
 \end{equation}
  where $c_n$ is the overlap of the initial starting state with that of the n-th eigenstate. 
  
  \section{Resource Scaling Calculations}
  
  In this section we count how many two-qubit gates will be needed for each qubit in order 
  to simulate one Trotter step of the time-evolution for several different plaquette Hamiltonians, 
  using the circuit identity (\ref{eq:qgate2}). For simplicity in these calculations, we are 
  considering only those qubits that correspond to physical links. The gate counting for 
  ancillary qubit(s) will be different, but the gate number will change across the different 
  Hamiltonians similarly to the way the physical-link-qubit gate counting will. Additionally 
  (\ref{eq:qgate2}) makes obvious that every two-qubit gate connected to an ancillary gate will 
  also already be counted by counting two-qubit gates that are connected to physical-link-qubits.
  
  Because the Hamiltonians we are discussing are always composed of plaquettes, and we are counting 
  how many two-qubit gates are necessary per link, we first need to determine how many plaquettes 
  touch a link as a function of the spatial dimension $d$.
  
  We can do this by first considering a particular link $l$ and a point in it $x$, and noting that 
  since there are $2 d$ links that touch every point in a square lattice, there are thus $2 d - 1$ 
  links other than $l$ that touch $x$. Each of these links--excluding the link that is co-linear 
  with the chosen $l$--will then correspond to a unique plaquette that touches $l$. There are thus 
  $2d-2$ plaquettes that touch each link in the lattice.
  
  \subsection{Quantum Link Models}
  
  For the $\mathbb{Z}_2$ quantum link model, we know from equations (\ref{z2a}) and (\ref{z2b}) that every plaquette has a single term corresponding to it that is a product of four Pauli-Z matrices. 
  From Appendix A, we have seen that we can write the exponential of this product in the form of 
  (\ref{eq:qgate2}) where $N=4$. Thus one plaquette product term of four Pauli-Z matrices corresponds 
  to two 2-qubit gates per physical-link-qubit. Thus there are $2(2d-2)$ two-qubit gates need for each 
  physical-link-qubit, using the plaquette-per-link number computed above.
  
  For the $U(1)$ quantum link model, we know from equation (\ref{explicitu1}) that every plaquette has 
  eight terms corresponding to it that are products of four Pauli matrices. This time they are not all 
  Pauli-Z matrices, however rotation is an operation that necessitates only single-qubit gates, so it 
  does not affect the two-qubit gate counting. Thus there will simply be eight times as many two-qubit 
  gates per physical-link-qubit as were needed for the $\mathbb{Z}_2$ model, and so the counting is 
  $8\cdot 2 (2d-2) = 16(2d-2)$ for the $U(1)$ quantum link model.
  
  For the $SO(3)$ quantum link model, we can realize the symmetry by defining two \textit{rishons} per 
  link, and then each plaquette consists of eight rishons. The terms in the Hamiltonian corresponding 
  to one plaquette are then
  \begin{equation}
    H_{\Box} = -\left(\vec{\sigma}_R^{1} \cdot \vec{\sigma}_L^{2}\right) \left(\vec{\sigma}_R^{2} \cdot \vec{\sigma}_R^{3}\right) \left(\vec{\sigma}_L^{3} \cdot \vec{\sigma}_R^{4}\right) \left(\vec{\sigma}_L^{4} \cdot \vec{\sigma}_L^{1}\right),
\end{equation}
 where $1,2,3,4$ correspond to links in a plaquette, and $R,L$ correspond to the two rishons in each link. 
 The dot products are over the three Paul matrix directions $x,y,z$, so there are $3^4=81$ terms for every 
 plaquette. The counting for each rishon-qubit (for a Trotter step) will thus be 81 times what is was for 
 the $\mathbb{Z}_2$ case, and thus it will be $81\cdot 2(2d-2)= 162(2d-2)$.

\subsection{Kogut-Susskind Model}
 We next consider the counting for the potential energy (plaquette terms) of the Kogut-Susskind model 
 \cite{PhysRevD.11.395}, which can be written as for Abelian gauge groups:
\begin{equation}
    V_{\mathrm{KS}} = -\alpha\sum_{\Box} \left(U_1 U_2 U^\dagger_3 U^\dagger_4 + {\mathrm{h.c.}}\right).
\end{equation}
 These $U$ operators are infinite dimensional in the full Hilbert space, but can be truncated to finite 
 representations in order to obtain finite Hilbert space formulations amenable to quantum simulation.

 For the $\mathbb{Z}_2$ theory, the smallest spin truncation possible yields $U_i = \sigma^z_i$, and 
 the Hamiltonian is exactly the same as the QLM Hamiltonian. Thus as before we will need $2(2d-2)$ 
 two-qubit gates for each physical-link-qubit.
 
 For the $U(1)$ theory, the smallest spin representation possible is spin-1. We can then write the $U(1)$ 
 theory in terms of $U_i (U^\dagger_i)$, which is the raising (lowering) operator for the electric fluxes. 
 In order to represent these three-state spins using qubits, we need two qubits per spin, which can be 
 represented as \cite{PhysRevD.103.114505}
 \begin{equation}
 \begin{aligned}
     U^x_i &= {\sigma_i^x}^{(1)} \otimes \left({\mathbbm{1}_i}^{(2)} + {\sigma_i^x}^{(2)} + {\sigma_i^z}^{(2)}\right)/2\\
     &\qquad + {\sigma_i^y}^{(1)} \otimes {\sigma_i^y}^{(2)}/2,\\
     U_i^y &= -{\sigma_i^y}^{(1)} \otimes \left(\mathbbm{1}_i^{(2)} +{\sigma_i^z}^{(2)} - {\sigma_i^x}^{(2)}\right) / 2\\
     &\qquad - {\sigma_i^x}^{(1)} \otimes {\sigma_i^y}^{(2)} / 2.
     \end{aligned}
     \label{xyu1}
 \end{equation}
 From (\ref{explicitu1}), we know that each plaquette will involve eight products of $U_i^{x/y}$ operators. 
 Since each of these operators is itself a sum of four terms, there will be $2\cdot 8\cdot 4^4  = 2\cdot 2048$ 
 two-qubit gates needed per qubit to produce one Trotter step of time evolution of the plaquette term. 
 Combining this with the $(2d-2)$ number for the plaquettes that touch each link, we have that we need 
 $2\cdot 2048 (2d-2)$ two-qubit gates per physical-link-qubit.
 
 \subsection{Symanzik Improvement}
   \begin{figure}
      \centering
      \includegraphics[scale=0.3]{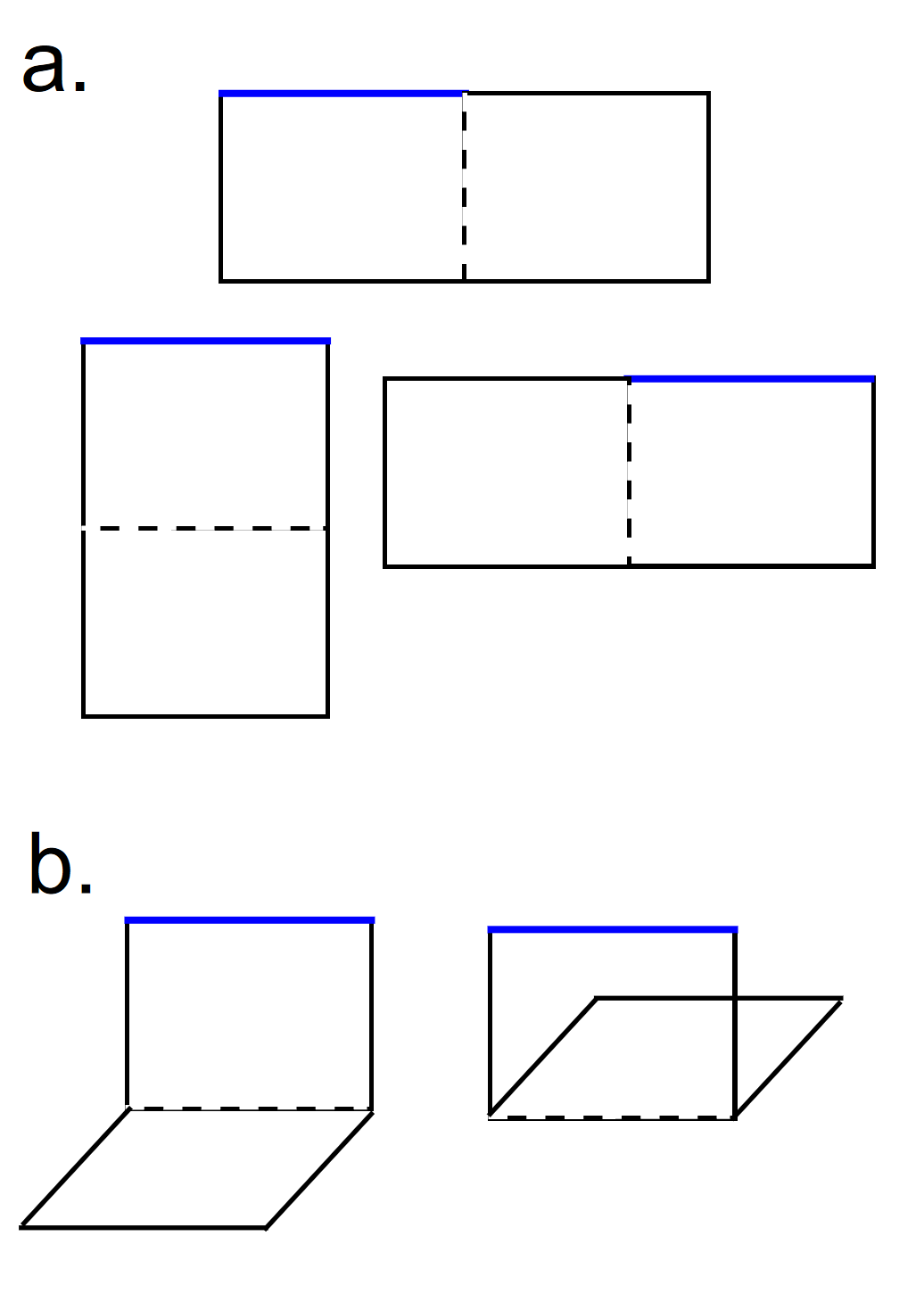}
      \caption{($a$) Rectangular loops that touch the blue link that can be formed from one plaquette that 
      touches the blue link. ($b$) Bent loops (for $d=2$) that touch the blue link that can be formed from 
      one plaquette that touches the blue link.}
      \label{fig:rectbent}
  \end{figure}
 Finally, we have that the Symanzik improvement \cite{SYMANZIK1983187} of the potential energy terms of the 
 Kogut-Wilson Hamiltonian involves the addition of the following terms \cite{Carena:2022kpg}:
 \begin{equation}
 \begin{aligned}
     V_{\mathrm{rect}} &= \alpha_1 \sum_{\mathrm{rect. \; loops}}  
     \left(U_1 U_2 U_3 U_4^\dagger U_5^\dagger U_6^\dagger + \mathrm{ h.c.}\right)\\
     V_{\mathrm{bent}}&= \alpha_2 \sum_{\mathrm{bent \; loops}}  
     \left(U_1 U_2 U_3 U_4^\dagger U_5^\dagger U_6^\dagger + \mathrm{ h.c.}\right),
     \end{aligned}
 \end{equation}
 where the rectangular and bent loops are formed from two adjacent plaquettes (either within the same plane 
 for the rectangular loops or in perpendicular planes for the bent loops). Diagrams illustrating these loops 
 can be found in Fig. \ref{fig:rectbent} \cite{Carena:2022kpg}.
 
 Just as we did for the plaquettes, we first need to determine how many rectangular and bent loops touch each 
 link. The rectangular loop can be viewed as a longer plaquette, and for every square plaquette that touches 
 a link, there are three corresponding rectangular loops that touch that link (see Figure \ref{fig:rectbent}), 
 because there are three sides of the first square plaquette that the second square plaquette can connect to. 
 Thus there are $3(2d-2)$ rectangular loops that touch each linking, using the plaquette number from before. 
 For the bent loops, there are $(2(d-1)-2)$ bent loops for every plaquette that touches a particular link 
 (see Figure \ref{fig:rectbent}), and this can be seen from using the previous formula of $2d-2$ (for number 
 of plaquettes that touch a link), to determine the number of plaquettes that touch a link of the first plaquette 
 that are perpendicular to the first plaquette. Because they must be perpendicular to the first plaquette, 
 we lose a dimension and the number is $2(d-1)-2 = 2d-4$. Thus the counting of bent plaquettes that touch 
 a particular link is $(2d-4)(2d-2)$. 
 
 For the $\mathbb{Z}_2$ gauge theory, we recall that we used the representation where $U_i=\sigma_i^z$. 
 From the numbers we just determined for the numbers of rectangular and bent plaquettes that touch a link, 
 we can conclude that we need $2\cdot 3(2d-2) + 2\cdot (2d-4) (2d-2)$  two-qubit gates per link-qubit per 
 Trotter step.
 
 For the $U(1)$ gauge theory, we know that $U_i$ can be written as $U^+_i$, and then writing in terms of 
 $U^x_i$ and $U^y_i$ will lead to $2^6/2 =32$ terms (similar to how \ref{explicitu1} has $2^4/2=8$ terms). 
 Because each $U^x_i$ and $U^y_i$ operator consists of four terms (\ref{xyu1}), there are 
 $32\cdot 4^6=32\cdot 4096$ products of Pauli matrices per rectangular/bent loop. Thus there are 
 $2\cdot 32\cdot 4096 \cdot 3(2d-2)$ two-qubit gates per link needed for the rectangular loops, and 
 $2\cdot 32\cdot 4096\;(2d-4)(2d-2)$ two-qubit gates per link needed for the bent loops, for one Trotter 
 step of evolution.
 
 \section{Fluctuations of IBM Q measurements}
 
  \begin{figure}
     \centering
     \includegraphics[scale=0.4]{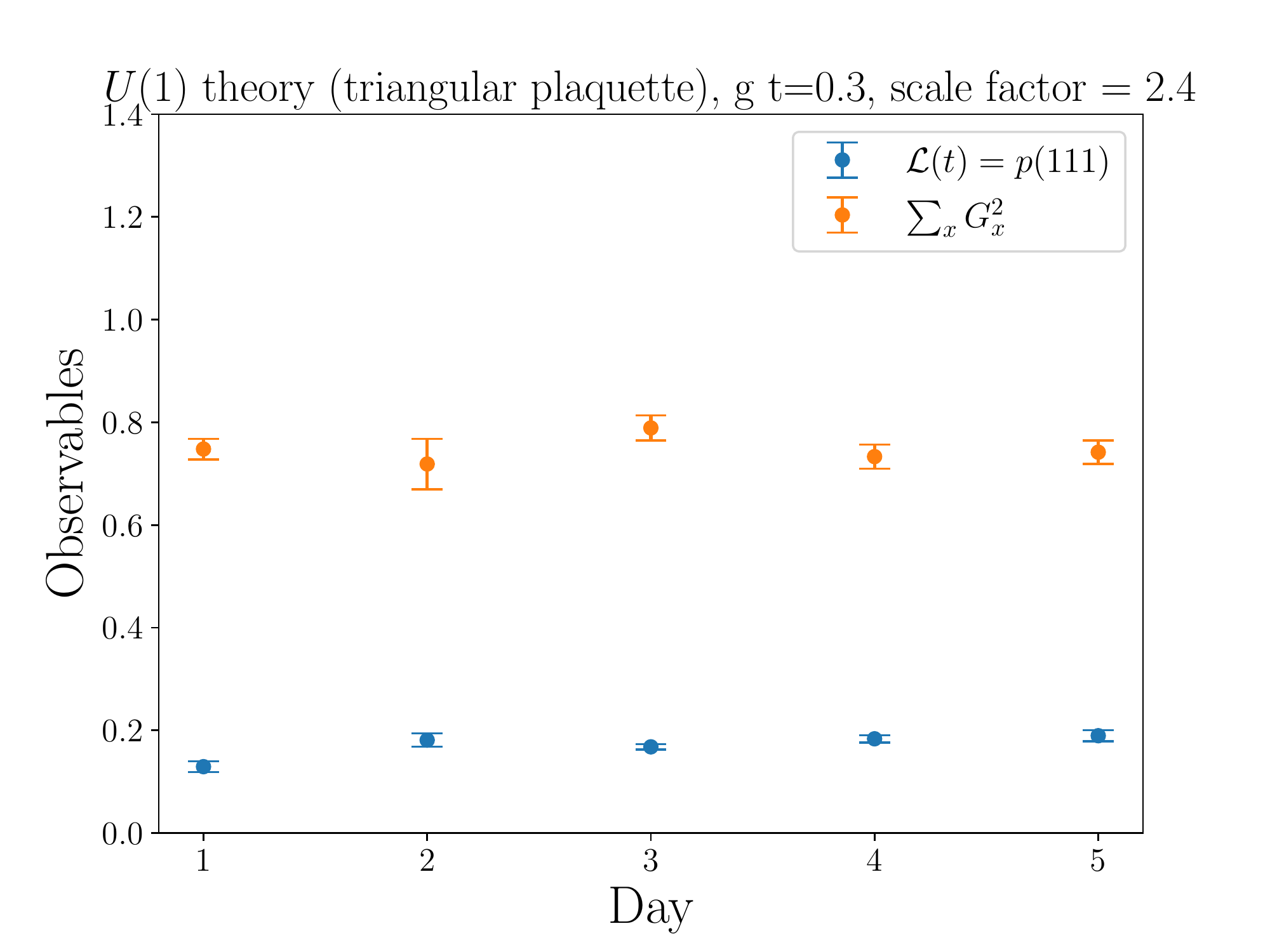}
     \caption{Data collected using IBM Q Quito over five consecutive days. The values 
     for $\mathcal{L}(t) = p(111)$ are $0.13(1)$, $0.18(1)$, $0.168(5)$, $0.183(7)$, and $0.19(1)$. 
     The values for $\sum_x G_x^2$ are $0.75(2)$, $0.72(5)$, $0.79(2)$, $0.73(2)$, and $0.74(2)$.}
     \label{dataovertime}
 \end{figure}
 To illustrate how measurement values can change from one calibration to the next, we did 
 the same measurements for the $U(1)$-theory on a triangular plaquette at a specific time and 
 folding scale factor ($g t=0.3$, scale factor$=2.4$) over five consecutive days. The values 
 we get for each day for two observables are plotted in Figure \ref{dataovertime}, and the 
 observable values are written directly in the caption. We computed the average and error for 
 each value of the observables by running the circuit for 8192 shots and five times within the 
 same day. The measurements have also been corrected for readout error.
 
 From the data we see that the measurements can vary substantially from each other from one calibration to the next. 
 The largest difference in this example is between the measurements for $\mathcal{L}(t) = p(111)$ 
 between Day 1 and Day 5, with the Day 5 measurement nearly $50 \% $ larger than the Day 1 measurement. 
 In taking our measurements and doing ZNE extrapolation for the figures in the main text, we made sure to take all scale factor 
 measurements for a particular data point on the same day, so that the extrapolation would make 
 sense for the noise of that day, but different time data for a time-evolution may come from different 
 days, as for each individual model it took 1-2 weeks to run these jobs on the IBM Q hardware for the real time evolution of the 
 observables over $g t\in [0,6]$ on a plaquette. While it would have been an improvement to then do an 
 additional average over 5-10 calibration days, the time to run this would have been prohibitive with 
 current queue wait time, and not indicative of what is straightforwardly achievable with this NISQ hardware.

\end{document}